\documentclass[aps,pra,twocolumn,twoside,floatfix,a4,showpacs,superscriptaddress]{revtex4}

\usepackage{float}
\usepackage{caption}
\usepackage{bbm}
\usepackage{amsmath, amssymb}
\usepackage{color}
\usepackage{graphicx,epsfig}
\usepackage{times} 
\usepackage{braket}
\usepackage{amsfonts}
\usepackage{amsthm}
\usepackage{amsmath}
\usepackage{multirow}
\usepackage[mathscr]{euscript}
\usepackage[stable]{footmisc}
\usepackage{MnSymbol}
\usepackage{bbold}
\usepackage{stmaryrd}
\usepackage{bbm}
\newfloat{graphic}{tbp}{lgr}
\floatname{graphic}{Graphic}

 \newcommand{\probfont}[1]{\mathbb{#1}} 

\begin{document}

\title{\textbf{Efficient Decoders for Qudit Topological Codes}}

\author{{Hussain Anwar\footnote{{These authors contributed equally to this paper.\label{astlabel}}}}}
\email{hussain.anwar@brunel.ac.uk.}
\affiliation{Department of Mathematical Sciences, Brunel University, Uxbridge, Middlesex UB8 3PH, UK.}
\affiliation{Department of Physics and Astronomy, University College London, Gower Street, London WC1E 6BT, UK.}
\author{Benjamin J. Brown\textsuperscript{\ref{astlabel}}}
\affiliation{Quantum Optics and Laser Science, Blackett Laboratory, Imperial College London, London SW7 2AZ, UK.
}
\author{Earl T. Campbell}
\affiliation{Dahlem Center for Complex Quantum Systems, Freie Universit{\"a}t Berlin, 14195 Berlin, Germany.}
\author{Dan E. Browne}
\affiliation{Department of Physics and Astronomy, University College London, Gower Street, London WC1E 6BT, UK.}
  
\begin{abstract}
Qudit toric codes are a natural higher-dimensional generalization of the well-studied qubit toric code. However standard methods for error correction of the qubit toric code are not applicable to them. Novel decoders are needed. In this paper we introduce two renormalization group decoders for qudit codes and analyze their error correction thresholds and efficiency. The first decoder is a generalization of a ``hard-decisions'' decoder due to Bravyi and Haah [arXiv:1112.3252]. We modify this decoder to overcome a percolation effect which limits its threshold performance for high $d$. The second decoder is a generalization of a ``soft-decisions'' decoder due to Poulin and Duclos-Cianci [Phys. Rev. Lett. {104}, 050504 (2010)], with a small cell size to optimize the efficiency of implementation in the high dimensional case. In each case, we estimate thresholds for the uncorrelated bit-flip error model and provide a comparative analysis of the performance of both these approaches to error correction of qudit toric codes.

\end{abstract}
\maketitle

%-----Sections-------------------------------------
\section{Introduction}\label{SecIntro}
The study of quantum error correction and fault-tolerant quantum computation \cite{S95,CS96,S96} for qubit systems is very well established, and the combination of topological codes \cite{K03} for robust error tolerance and magic state distillation \cite{BK05} for universality has become a leading framework for fault-tolerant quantum computation \cite{RHG07,FG09, RH07,WFH11}.

In contrast to qubit systems, fault-tolerant quantum computation with systems of dimension $d$ higher than 2 are less well understood. Only recently were the first magic state distillation schemes for qudit systems developed, which demonstrated improved distillation thresholds and reduced overhead compared with their qubit counterparts \cite{ACB12,CAB12}. Also, recently the first decoders for qudit topological codes were proposed \cite{CP13}, providing the ingredients for a fault-tolerant qudit computation.  

\begin{figure}
  \centering
    \vspace{-.2cm}
    \includegraphics[width=0.2\textwidth]{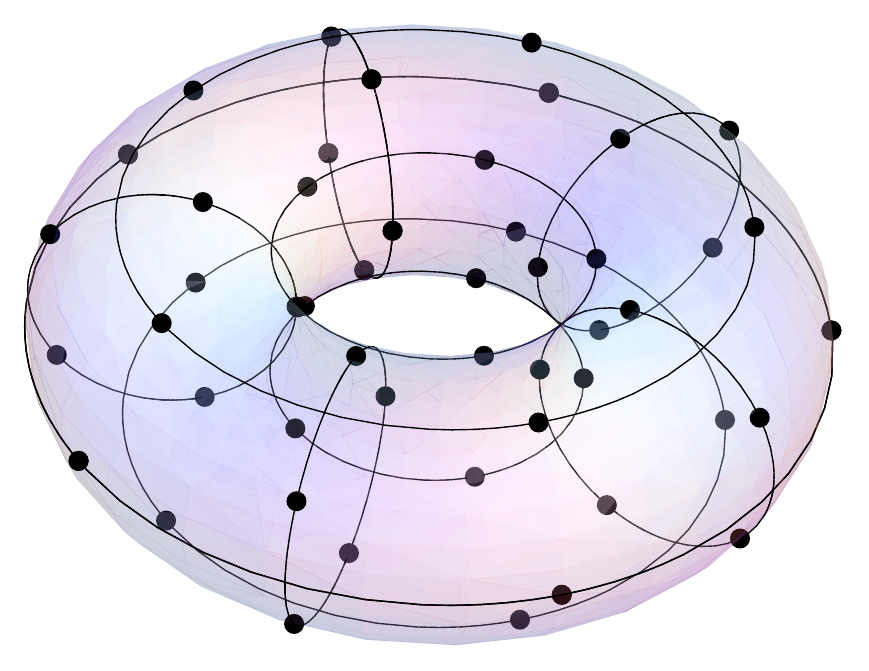}
    	\captionsetup{singlelinecheck=yes, justification=centerlast}
    	\caption{In a toric code, qudits (black dots) are attached to the edges (black lines)  of a square lattice on the surface of a torus. Errors are detected by measuring the code's stabilizer generators---operators which act locally on four qudits associated with each plaquette and vertex. }
    	 \label{Torus}
     	\vspace{-.3cm}
\end{figure}

Topological codes were introduced by Kitaev in one of his seminal papers \cite{K03}, establishing a framework encompassing both qubit and qudit variants.  In particular, the toric code places qudits on the surface of a torus, as illustrated in Fig.~\ref{Torus}. Also notable is the planar code, which has similar properties but can be physically realized in two-dimensions \cite{BK98,FM01}. These codes are topological in nature as the quantum information is encoded in degrees of freedom which are independent of the local physics of the code.  

For a topological code to protect quantum information, physical errors must be detected and corrected at a sufficient rate to prevent the errors from accumulating and causing an unwanted logical error. An important part of error correction is the \textit{decoder}, the classical algorithm which, given the output of the error detecting measurements (the \textit{syndrome set}) computes a correction operator which should restore the quantum code to its original state. The mapping between syndromes and errors can never be one-to-one (even in classical codes), so a good decoder will output a correction operator which has a high likelihood of successful error correction. Comparing decoders, however, is subtle. Whilst some decoders may achieve the highest thresholds (\textit{optimal} decoders) others may run faster and at more favorable computational cost.

In this work, we implement and refine two types of decoders for the qudit generalization of Kitaev's toric code.  One of the motivations behind exploring qudit systems is that stabilizer measurements have more outcomes, and so provide more information with which to determine the error locations. If this extra information is exploited correctly, then improvement in performance vis-\`{a}-vis qubit codes can be observed \cite{AMT12}. Studies of the thermal stability of the toric code also indicate some advantages in using qudit systems \cite{VRD12}. Moreover, the extra levels in qudit systems allow for the enhancement of quantum memories in two-dimensions via the insertion of domain walls \cite{BSW11, KK12}, which is not possible in qubit codes \cite{BAP13}.

For the \textit{qubit} toric code a variety of decoding algorithms have been studied. 
The early extensive study by Dennis \textit{et al.} \cite{DKL02} demonstrated how the decoding problem for the qubit toric code undergoing independent $X$ and $Z$ errors can be mapped to a statistical mechanical model, the random-bond Ising model (RBIM). Remarkably, they showed that the optimal error threshold, $ p_\mathrm{th}^\mathrm{opt} $, for this model directly maps to a phase transition point in the RBIM, known as the Nishimori point \cite{NishimoriBook}, which had already been identified numerically to be around $ 10.9\% $ \cite{HPP01,MC02,O09,Q09}. Moreover, Dennis \textit{et al.} observed that this optimal threshold lay very close to an important quantity arising in quantum Shannon theory called the \textit{hashing bound threshold} $ p_{\mathrm{th}}^\mathrm{H}$ \cite{BDS96} (see App. \ref{AppSecHash}). In the qubit case, and for the independent $X$ and $Z$ error model, the hashing bound threshold is $ p_{\mathrm{th}}^\mathrm{H}=11.0028\% $.  Further work by Takeda and Nishimori \cite{TN05} showed that the similarity between the optimal and the hashing bound thresholds applies to more general statistical mechanical models. They made a conjecture in statistical mechanics terms, which when translated to topological codes, implies that the optimal code threshold should coincide with the hashing bound threshold for a variety of error models, i.e. $ p_{\mathrm{th}}^{\mathrm{opt}}=p_{\mathrm{th}}^{\mathrm{H}}$. Whether this conjecture holds (either exactly or approximately) remains an open question, but numerical results so far have supported it \cite{BAO12}.

Thresholds are dependent on the error model chosen. In this paper, unless explicitly stated otherwise, all thresholds reported are for the independent $X$ and $Z$ error model, defined in Sec. \ref{SecNoise}. Note also that the reported threshold values are code thresholds and not fault-tolerance thresholds and that error-free syndrome measurement and correction is assumed.

The most typically employed decoder for the qubit toric code is the minimum-weight perfect matching algorithm (MWPMA), which has an efficient implementation based on Edmonds' blossom algorithm \cite{E65}. This algorithm has been extensively studied and its threshold has been estimated to be $ 10.3\% $  \cite{WHP03}. As we will see below, however, the MWPMA is not suitable for decoding $d>2$ qudit toric codes.

Recently, a number of alternative decoders have been proposed. Of particular relevance to this paper are the two decoders which utilize renormalization group (RG) ideas, proposed by Duclos-Cianci and Poulin \cite{CP10,CP102}, and by Bravyi and Haah \cite{BH11}. While both of these decoders employ RG techniques, they do so in very different ways. To distinguish between them, we shall use the terminology suggested by Duclos-Cianci and Poulin  \cite{CP13b}. We will refer to the first as the \textit{soft-decisions}-RG (SDRG)  decoder and the second as the \textit{hard-decisions}-RG (HDRG) decoder. These decoders will be detailed in Secs. \ref{SecRG1} and \ref{SecRG2}, and the reasons for this terminology will become clear. These decoders have gained great attention recently due to the fact that they have a greater flexibility over MWPMA \cite{SR12,HDCP12} and yet achieve comparable thresholds.  The thresholds that were obtained for the qubit toric code by the SDRG and HDRG decoders are $ 8.2\% $ and $ 6.7 \% $, respectively.

In Sec. \ref{SecRG1} we generalize the HDRG decoder to decode the qudit toric code. We will show that the construction of this decoder has no dependence on the qudit dimension, which allows us to obtain a numerical estimate of the threshold for any dimension. Some of the refinements of our qudit decoder also have benefits for the qubit code. We will demonstrate that the original qubit HDRG decoder of \cite{BH11} can be improved so that a higher threshold of $ 8.4\% $ is achieved. In the limit of high dimensions this decoder reaches a saturating threshold of about $ 18\% $. We discover that this behavior is due to a percolation effect, where thresholds achieved by this decoder are always upper-bounded by the \textit{syndrome} percolation threshold. To beat this upper-bound, we introduce an \textit{initialization step} into the algorithm which disrupts this percolation effect and enhances the performance of the decoder such that, for high dimensional systems, thresholds as high as $ 30 \% $ are obtained. We refer to the HDRG decoder when augmented with the initialization step as the {\em enhanced}-HDRG decoder. 

\begin{figure}
  \centering
    \includegraphics[width=0.5\textwidth]{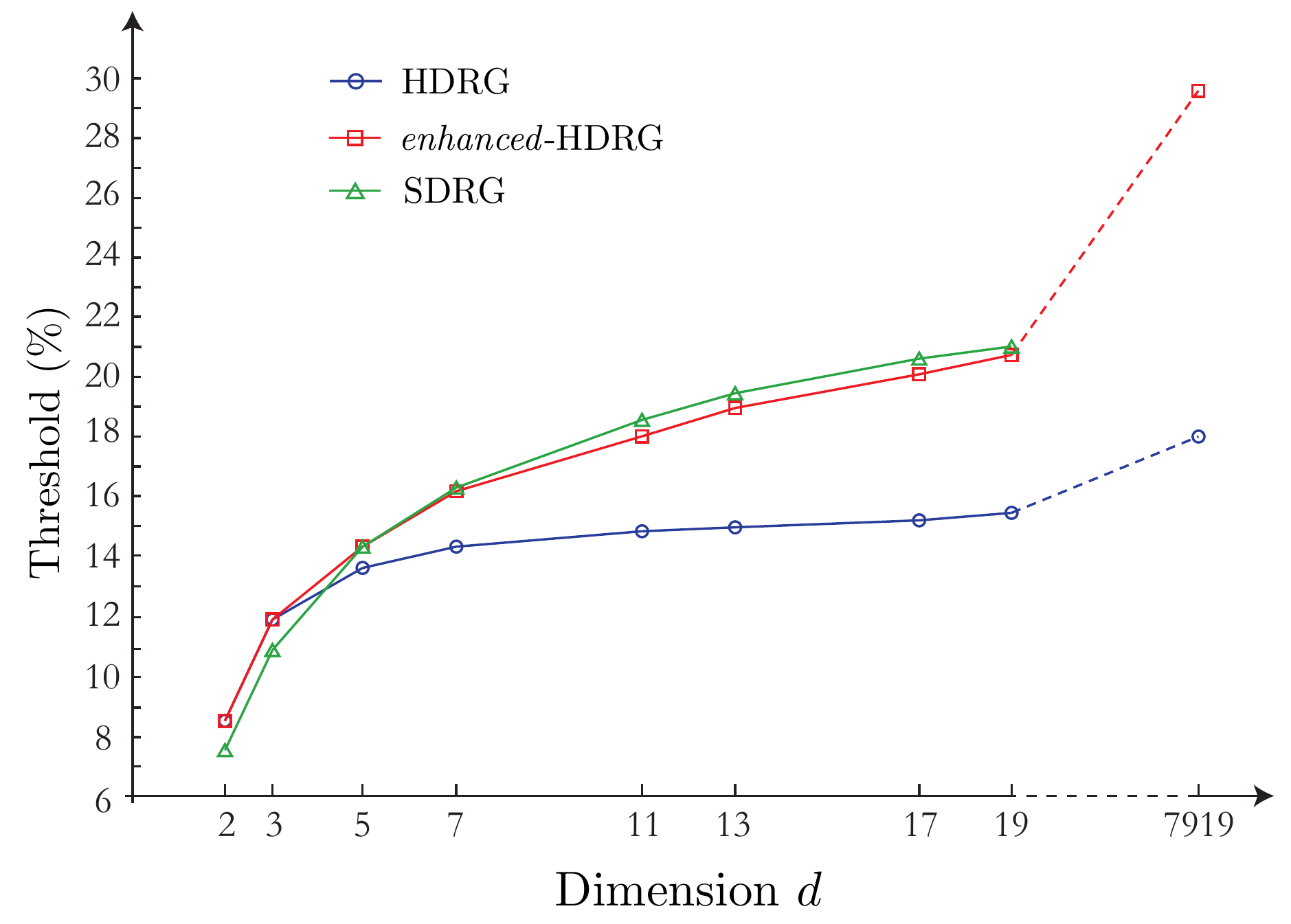}
      \captionsetup{singlelinecheck=yes, justification=centerlast}
      \caption{A comparison between the thresholds obtained using the HDRG, enhanced-HDRG and SDRG decoders presented in this paper. The dimension $ 7919 $ is the $ 1000 $th prime number used here to demonstrate the independence of the HDRG decoder on the dimension of the qudits. The error bars has been omitted for clarity.}
      \label{Comparison}
      \vspace{-.5cm}
\end{figure}

In Sec. \ref{SecRG2} we develop a variant of the SDRG decoder for toric codes of any dimension. Recently, Duclos-Cianci and Poulin generalized the SDRG decoder for qudit codes \cite{CP13}. They studied the relationship between the hashing bound threshold and the thresholds achieved by their decoder,  finding, strikingly, that their thresholds approximate a constant fraction of the hashing bound threshold. Due to the fact that their decoder has a run-time complexity that depends on the dimension as $ O(d^{7}) $, they were not able to investigate this behavior beyond $ d=6 $. The version of SDRG encoder that we develop has a different scaling behavior with a dimension dependence of $O(d^{4}) $. The smaller cell size we use lead to lower thresholds, but allow us to analyze much higher dimensional systems. This enables us to estimate thresholds for systems of dimensions up to $d= 19 $. 

We have summarized the thresholds obtained by our HDRG, enhanced-HDRG and SDRG decoders in Fig.~\ref{Comparison}. This paper is structured as follows. In Sec. \ref{SecToric} we review the qudit toric code. In Sec. \ref{SecNoise} we describe the noise model formally and describe the numerical method we adopt to estimate the thresholds. Secs. \ref{SecRG1}  and \ref{SecRG2} describe our generalization of the HDRG and SDRG decoders, respectively. Finally, we compare and discuss these two decoders in Sec. \ref{SecSum}.

\section{Qudit Toric Code}\label{SecToric}

The toric code is a stabilizer error-correcting code described within the stabilizer formalism~\cite{G97phd,NC00}. Here, we consider a toric code consisting of a lattice of $d$-level quantum systems, {\em qudits}, where $ d\ge 2 $. For simplicity, we will restrict our discussion to prime dimensions. Our definitions are valid in  both the $d=2$ and $d>2$ case.

 The single qudit Pauli group, $ \mathcal{P}_d $, is generated by 
\begin{equation}\label{EqPauli}
X=\displaystyle\sum_{j\in\mathbb{Z}_{d}}\ket{j\oplus 1}\bra{j}, \quad Z=\displaystyle\sum_{j\in\mathbb{Z}_{d}}\omega^{j}\ket{j}\bra{j},
\end{equation} 
where the addition ``$ \oplus $" is carried out modulo $ d $, and $ \omega=e^{2\pi i/d} $ \cite{G99}. These operators satisfies the ``omega-commutation relation'' $ XZ=\omega^{-1}ZX $. 

In the case where qudit dimension $d>2$, unitaries $X$ and $Z$ are not Hermitian, but, possessing orthogonal eigenspaces, they still can be interpreted as observable measurements whose measurement outputs are labeled by complex eigenvalues $\omega^k$. As a short-hand we shall usually denote outcome $\omega^k$ simply by integer $k$.

The $ n- $qudit Pauli group $ \mathcal{P}^{n}_{d} $ is generated by the $ n- $fold tensor product of single qudit Pauli operators. 

A stabilizer code is defined by an Abelian subgroup $ \mathcal{S} $ of the Pauli group $ \mathcal{P}^{n}_{d} $. The common ``$+1$'' eigenspace of $ \mathcal{S} $ forms a protected subspace called the code-space, and the elements of $ \mathcal{S} $ are called the \textit{stabilizers} of the code. 

\begin{figure}
  \centering
    \includegraphics[width=0.5\textwidth]{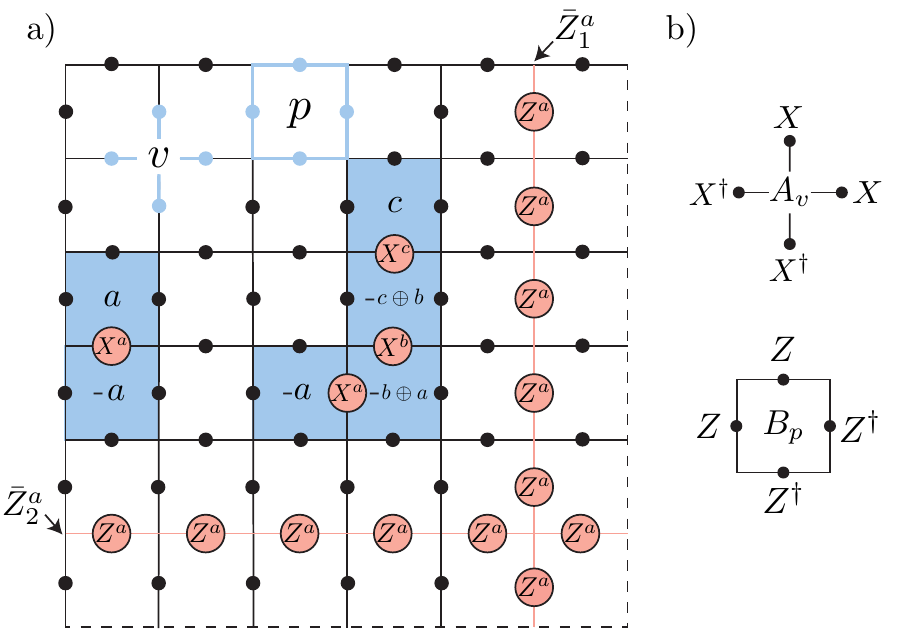}
      \captionsetup{singlelinecheck=yes, justification=centerlast}
      \caption{a) The primal lattice of the toric code with qudits depicted by black dotes, and a couple of examples showing how the plaquette syndromes are generated for a set of $ X $ errors. The two logical operators $ \bar{Z}_{1}^{a} $ and  $ \bar{Z}_{2}^{a} $  correspond to the two non-contractible loops on the primal lattice. The dashed lines indicate the periodic boundaries. b) The qudit vertex and plaquette operators. 
      }
      \label{FigLattice}
      \vspace{-.5cm}
\end{figure}

The qudit toric code is defined on a square lattice of $L\times L$ vertices with periodic boundary conditions, where we have a qudit on each of the edges of the lattice, see Fig.~\ref{FigLattice}(a). The stabilizer group of the qudit toric code is generated by two types of stabilizers, the {\em vertex} operators $A_v$ and the {\em plaquette} operators $B_p$. There is an $A_v$ stabilizer for each vertex $v$ and a $B_p$ stabilizer for each plaquette $p$ of the lattice. We show examples of $A_v$ and $B_p$ operators in Fig.~\ref{FigLattice}(b). 
The qudit toric code encodes two logical qudits. The operators acting within the code-space on the logical information are string-like operators that have support on non-contractible loops of the lattice. We show examples of logical operators $\bar{Z}^a_1$ and $\bar{Z}_2^a$ in Fig.~\ref{FigLattice}(a). The conjugate $\bar{X}_j^a$ logical operators have support over non-contractible loops on the \textit{dual} lattice. Logical operators commute with the stabilizer group, but are not members of $\mathcal{S}$. We shall label the initial encoded state, which we wish to preserve, as $\ket{\psi_\textrm{init}}$.

The vertex and plaquette operators collectively comprise the \textit{check} operators for the code. These are the operators which are measured to determine the error syndromes. We shall call the integer outcome $a$ of a measured check operator a \textit{syndrome}. If this outcome is zero we call this a trivial syndrome.  The combined outputs of all check operators we shall call the syndrome set.  The syndrome set thus comprises an integer $a$ for each vertex $v$ and plaquette $p$ check operator, such that $a = 0,\, 1,\dots,\,d-1$ correspond to the eigenvalues $\omega^a$ of the measured operator. 

After measuring the set of stabilizer generators, errors are projected to Pauli errors and the lattice will be in the state $E\ket{\psi_\textrm{init}}$ such that  $E  \in\mathcal{P}_d^n$. In common with other CSS codes \cite{CS96, S96}, we can deal with $X$-type and $Z$-type errors on the toric code independently. Here we will focus on $X$ errors, however the behavior of $Z$ errors is equivalent with respect to the dual lattice. The relationship between the errors and the syndromes is crucial to the design of the decoder, in essence the decoder is trying to invert this mapping. 

If a toric code state undergoes a single $X^a$ error, as shown for example in Fig.~\ref{FigLattice}(a), then a pair of non-zero syndromes are generated with values $a$ and $-a$, where negative numbers are defined modulo $d$. For multiple errors the integer syndromes combine \textit{additively}. For example, the three errors $X^a$, $X^b$ and $X^c$ in Fig.~\ref{FigLattice}(a) generate syndromes $-b\oplus a$ and $-c\oplus b$ on the plaquettes between pairs of errors. Note that in general, any set of syndrome outcomes generated by a set of errors will sum to zero (e.g. for a single error $a+(-a)=0$).

We see that if adjacent errors correspond to equal (in some directions) and opposite (in other directions) powers of $X$, then the intermediate syndrome cancels to the zero outcome. It is this behavior which strongly distinguishes qubit toric codes from $d>2$ qudit codes. For qubits, Pauli operators $X$ and $Z$ are self-inverse, which means that intermediate syndromes always cancel. A string of errors will only generate non-trivial syndromes at its ends. For $d>2$, however, the syndrome between adjacent errors will cancel only in the special case that adjacent errors are equal (or opposite, depending on the direction). This is the principle reason why decoders designed for the qubit code will often not immediately generalize to the $d>2$ setting.

The task of the decoder, then, is to derive a correction operator $F^\dagger$ which returns the state to the code-space.  Given any correction operator $F^\dagger$ which (trivially) returns the logical operator to the code-space, and given any logical operator on the code-space $L$, then the combined operator $L F^\dagger$ will also return the system to its code-space. Indeed, given any set of errors $E$ and any logical operator $L$, the error set $EL$ will return an identical syndrome. It is this degeneracy between the errors and the syndromes which makes decoding non-trivial. For a successful correction, $F$ must be chosen such that $F^\dagger E $ acts trivially on the encoded logical information, i.e. implements the identity operator and not a non-trivial logical operator.
 
Considering $X$ operators alone, the logical operators on the toric code are of the form $\bar{Z}_1^j \bar{Z}_2^k$ for all $ j $ and $ k $ integers between $ 0\le j,k \le d-1 $. There are thus $d^2$ equivalence classes of correction operators for any given syndrome. It is the role of the decoder to choose the most appropriate correction operator given the underlying error model. We call a decoder with the highest probability of choosing the correct equivalence class of correction operators the optimal decoder.

It is sometimes useful to adopt {\em quasi-particle} terminology to describe syndromes on a toric code. We interpret a single syndrome outcome $a$ as representing a quasi-particle of {\em charge} $a$, with $-a$ outcome representing the anti-particle of $a$. The set of quasi-particles generated by any set of errors is then of neutral charge, and sets of quasi-particles of overall neutral charge can be \textit{annihilated} by the application of an appropriate correction operator. 

The mathematical structure of the toric code can be elegantly and concisely described using the  mathematics of \textit{homology}.  Homology theory provides a concise and exact representation of the (otherwise somewhat complicated) relationship between errors and syndromes, and the logical operators on the code-space. Since we do not expect the typical reader of this paper to be familiar with this field, we have aimed to refrain from using homological terms, as much as we can, in the main text and provide a primer on homology in App.~\ref{HomologyAppendix}. Some aspects of SDRG decoder, however, are most cleanly expressed in homology terms and these shall be explained at the beginning of Sec.~\ref{SecRG2}.  A powerful aspect of homology is that the relationship between errors and syndromes on the toric code is the same for $d=2$ and all higher values of $d$. 

It is worth noting, we do not choose to generalize the MWPMA. In the qubit case, the MWPMA typically consists of constructing a complete weighted graph where the vertices are non-trivial syndromes and the weight of the edges is the shortest Manhattan distance between the vertices. Then using Edmonds' algorithm \cite{E65,K09} the perfect matching of minimum weight can be efficiently determined. In quasi-particle language MWPMA works by associating pairs of syndromes, which together have neutral charge. It then finds the minimal weight correction operators to annihilate each pair. In higher dimensions, however, the charge of the vertices ranges across the set $ \{1,\dots,d-1\} $. Any neutral set (not just a pairing) of clusters should be considered for annihilation.  Thus to find the lowest weight error correction chains for the qudit code requires an algorithm, which must minimize weights on a hypergraph whose hyperedges consist of all charge neutral subsets of vertices. Minimum weight hypergraph matching is in general an NP-Hard problem \cite{K72}. We therefore do not expect good performance for such a decoder, and have not pursued it here. 

\section{Noise Model and Threshold Estimation}\label{SecNoise}

In this section we define the noise model used in this paper, and review the numerical methods we use to estimate the thresholds. We work with the independent error model, otherwise known as the uncorrelated error model, which is widely used in other studies of the toric code \cite{DKL02,CP13}. The principle benefits of this model are its simplicity and its direct mapping to statistical mechanics models (e.g. the RBIM and Potts gauge glass models).

In the independent error model, we treat $X$-type and $Z$-type errors as separate processes which act independently on each physical qudit. Each channel has a simple definition. For $X$-type errors, with probability $1-p$ no error occurs to the qudit, and with probability $p/(d-1)$ the error operator $X^j$ is applied, where $j$ is an integer between $ 1<j\le d-1 $. The $Z$-type errors occur according to an analogous channel with the same probability $p$.  The important features of this error model is that all powers of $X$ occur with equal likelihood, and $X$ and $Z$ errors are uncorrelated.

We will estimate these thresholds numerically via a Monte Carlo simulation. For a single Monte Carlo sample, we initiate the lattice in the pure state of the code-space, fix $p$ and then generate a random error configuration using the above noise model. The syndromes of the error configuration are then measured and fed to the decoder. The decoding algorithm will return a Pauli correction operator $F^\dag$, that will return the system to the code-space.

In the simulation we repeat this procedure $ N $ times for a given $p$, and we evaluate the success probability $ p_\text{succ} $ as the fraction of times the decoder succeeds. The standard deviation in the estimated success probability is $ \sigma=\sqrt{p_\text{succ}(1-p_\text{succ})/N} $. To determine the threshold, we plot $p$ versus $p_\text{succ}$ for different lattice sizes as shown, for example, by Figs.~\ref{QubitPlot} and~\ref{QutritPlot}. The threshold $p_{\text{th}}$ is defined to be the \textit{point} at which the success probability curves intersect in the limit $ L\rightarrow\infty $. In other words, the threshold represents the point below which arbitrarily high $p_\text{succ}$ can be achieved provided that the lattice is made large enough. However, in the actual simulations, the data points can only be obtained for a relatively small lattice sizes $L$, and such lattices are subject to small system size effects, which can affect the evaluation of $p_{\text{th}}$. This is easily seen in the $L=16$ curve of Fig.~\ref{QubitPlot}. 

To account for the small system size effects, we estimate $p_{\text{th}}$ by using the fitting proposed by Harrington \textit{et al.} \cite{WHP03}. In this fitting, \textit{all} the data points are fitted to the curve 
\begin{equation} 
p_{\text{succ}}=A+Bx+Cx^2+DL^{-1/\mu}, \label{HarringtonFit}
\end{equation}
where $ x=(p-p_{th})L^{1/\nu} $, as shown, for example, in the boxed plot in Fig.~\ref{QubitPlot}. The last term in the fitting, $DL^{-1/\mu}$, accounts for the the small size effects. We can see that, in the limit of $L \rightarrow \infty$, this term tends to $ 0 $, where $\mu$ is positive \cite{NoteFitting}.

In the next two sections we introduce two types of decoders, namely, the HDRG and SDRG decoders, suitable for decoding the qudit toric code of any dimension. We describe earlier formulations of these decoders, our refinements to them and the resulting thresholds achieved for the independent error model.

\section{\em{hard-decisions}-RG Decoder}\label{SecRG1}

The HDRG decoder was introduced by Bravyi and Haah in \cite{BH11} as an efficient decoder for general local topological codes \cite{H11,BH13}, where MWPMA is an unsuitable decoder. For the qubit toric code, they obtain a threshold of  $6.7\%$ using the independent noise model. The construction of this decoder built upon previous work by Harrington \cite{H04} and Dennis \cite{D05}.  In the first sub-section of this section, we present a refined version of this decoder and show that these refinements achieve higher thresholds for the toric code.

\subsection{Decoder Description}\label{SecHDRG}

The HDRG decoder has a simple and elegant intuition behind its construction, and before we introduce it formally we shall give a heuristic description of how it works. If error rates are low, errors will typically be sparsely distributed. Each cluster of errors will generate a cluster of non-zero syndrome measurements in its immediate vicinity. Identifying and annihilating such small clusters of syndromes is the heart of Bravyi and Haah's HDRG decoding technique. 

The HDRG decoder is executed with run-time complexity of $O(L^2)$. The decoder considers the syndromes on the lattice over many levels of decoding. Each level of decoding is associated with a geometric measure of distance on the square lattice, such that the distance gets bigger as the levels increase. At each level, the non-trivial syndromes are divided into clusters which are disjoint with respect to this measure. In other words, in each cluster, all of the syndromes are separated from at least one other syndrome of the cluster by a distance no greater than that determined by the decoding level. If the sum of all the syndromes within a cluster is zero (modulo $ d $), i.e. if the cluster is neutral, then the syndromes of the clusters are annihilated locally with respect to the cluster. Clusters whose total syndromes is non-zero are referred to as \textit{charged} clusters. Charged clusters are passed to the next level until ultimately they become part of a neutral cluster which is annihilated. Next, we will define and explain all the aspects of this decoder more rigorously.

\begin{figure}
  \centering
    \includegraphics[width=0.4\textwidth]{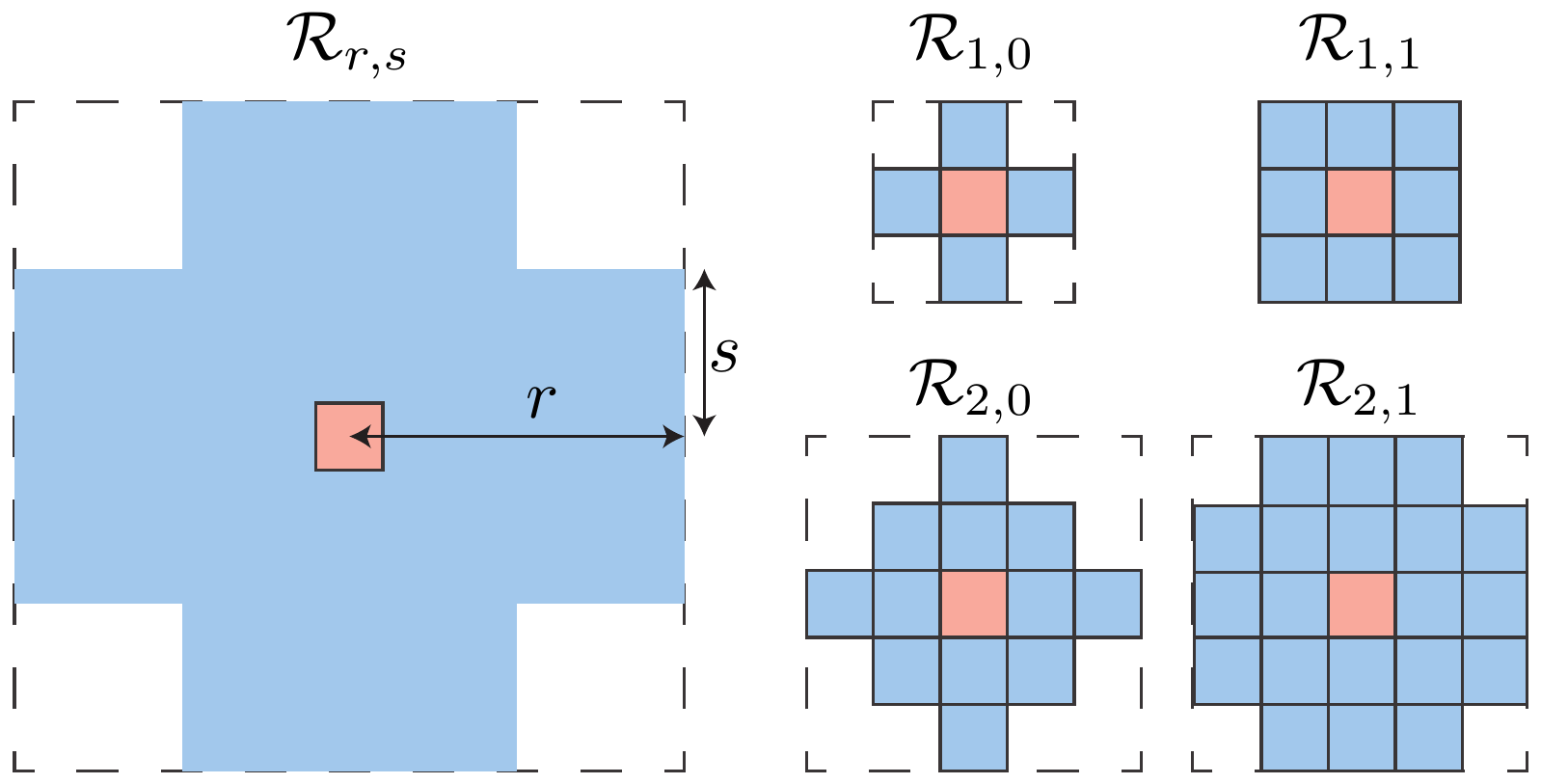}
    \hspace{-5mm}
      \captionsetup{singlelinecheck=yes, justification=centerlast}
      \caption{The refined regions $ \mathcal{R}_{r,s} $ on a taxi-cab geometry (left hand side) with examples of the first four levels (right hand side).}
      \label{DistanceMeasure}
      \vspace{-.5cm}
\end{figure}

Without loss of generality we shall consider only the plaquette syndromes, with the analogous vertex formalism being obvious. Firstly, we review some distance measures between plaquettes labeled as $\vec{x}=( x_{1}, x_{2})$ and $ \vec{y}=(y_{1}, y_{2})$.  The Manhattan (or taxi-cab) distance is $D^{(1)}( \vec{x}, \vec{y}) =  |x_{1}-y_{1}| + |x_{2}-y_{2}|   $.  Our second distance is the Max distance, and is $D^{(\infty)}(\vec{x},\vec{y}) = \max \{ | x_{1}-y_{1}| , | x_{2}-y_{2}| \}$.   Both can be used to define balls of a certain radius, centered on a plaquettes $p$, such that
\begin{eqnarray}
    B^{(1)}_{r}( \vec{x}) = \{ \vec{y} | D^{1}( \vec{x}, \vec{y}) \leq r \} , \\ 
    B^{(\infty)}_{r}( \vec{x}) = \{ \vec{y} | D^{\infty}(\vec{x}, \vec{y}) \leq r \}.
\end{eqnarray}
Although called balls, the Max distance generates squares and the taxi-cab distance picks out diamonds.  However, primarily we are interested in regions that combine these two regions.  For any integers $r$ and $s$, we define regions $\mathcal{R}_{r, s}$ that when centered on a point $\vec{x}$ are
\begin{equation}
     \mathcal{R}_{r, s}(\vec{x}) =  B^{(1)}_{r+s}(\vec{x})  \cap B^{(\infty)}_{r}(\vec{x}) ,
\end{equation}
and so simply the intersection of two balls with different metrics.  The first few instances are shown in Fig.~\ref{DistanceMeasure}, and clearly we only need to consider $s \leq r$.  Note that the regions are symmetric, so if $\vec{x} \in \mathcal{R}_{r, s}( \vec{y} )$ then $\vec{y} \in \mathcal{R}_{r,s}( \vec{x} )$ and when this happens we say $\vec{x}$ and $\vec{y}$ are $(r,s)$--connected.

Furthermore, we need a notion of connection for a \textit{cluster} $ C $, or set, of plaquettes.  Firstly, we define connected paths in $C$.  A path $\gamma$ in $C$ is an \textit{ordered} subset of $C$,  such that $\gamma=\{\vec{x}^{(1)}, \vec{x}^{(2)},\dots,\vec{x}^{(n+1)}\}$, and we define it to be an $(r,s)$--path if for all $j$, $\vec{x}^{(j)}$ and $\vec{x}^{(j+1)}$ are $(r,s)$--connected.  Now, we say the cluster is an $(r, s)$--cluster if for all $\vec{x}, \vec{y} \in C$ there exists an $(r,s)$--path in $C$ starting at $\vec{x}$ and ending at $\vec{y}$. The intuition behind considering these particular regions is to take into account some of the degeneracy in the errors creating the syndromes. We will discuss this point in more details in the next section.

These geometric concepts can be used to explain the decoding scheme.  Given the measurement data of all the plaquettes, if plaquette $\vec{x}$ has measurement outcome $m_{\vec{x}}$, then information is conveyed by the ordered pair $(\vec{x}, m_{\vec{x}})$, and the full list of charged plaquettes is $\mathcal{W}= \{ (\vec{x}, m_{\vec{x}}) | m_{\vec{x}} \neq 1) \}$.  Similarly, a charged cluster is a subset of the full charge distribution, $\mathcal{C} \subset \mathcal{W}$, where we use a different script to indicate the presence of charge information.  A charged cluster is said to be \textit{neutral} if the \textit{total} charge is zero, so that $\sum_{\mathcal{C}}m_{\vec{x}} =0$ modulo $d$.  Neutral clusters can always be annihilated by transporting and fusing the syndromes within the cluster until the total charge disappears.  When doing so,  we update the plaquette information from $\mathcal{W}$ to $\mathcal{W}'$, such that the annihilated neutral cluster $\mathcal{C}$ is no longer contained in $\mathcal{W}'$.  Despite there being many different ways to annihilate the charge, all of these are equivalent assuming only that the cluster is small compared to the lattice and the charges are transported within the cluster. The intuition behind the HDRG decoder is that if such small clusters are annihilated locally, then the resultant correction chains, combined with the actual error chain, will form a trivial loop of errors. That is, a stabilizer element of the code and so equivalent to the identity on the code-space. %Topologically, such chains are homologically trivial loops as we saw in the last section.  

The complete set $\mathcal{W}$ can always be partitioned into a set of \textit{disjoint} clusters $\tilde{\mathcal{W}} =\{ \mathcal{C}_1,\mathcal{C}_{2},\dots, \mathcal{C}_{n} \}$, for some $n$, and where $\mathcal{W}=\mathcal{C}_1\cup\mathcal{C}_{2}\cup\dots\cup\mathcal{C}_{n}$. We say a particular partition $\tilde{\mathcal{W}}$ is a $(r,s)$--partition if both the following conditions are satisfied: 
\begin{enumerate}
	\item[i.]  \textit{connectivity:} every charged cluster in the partition is an $(r,s)$--cluster;  
	\item[ii.] \textit{maximality:} for any distinct pair of charged clusters in the partition, $\mathcal{C}_{j}$ and $\mathcal{C}_{k \neq j}$, we find that $\mathcal{C}_{j} \cup \mathcal{C}_{k}$ is not $(r,s)$--cluster.
\end{enumerate}
Maximality tells us there is no suitable path between the disjoint clusters, and so they could not be merged into a single cluster.  Furthermore, whenever the connectivity condition is met, but maximally fails, there exists another partition that does fulfill both conditions using fewer charged clusters.

As stated previously, the HDRG decoder involve multiple levels of decoding. Each decoding level $ l $ is associated with a choice of regions $ \mathcal{R}_{r,s}$. At the first level we begin with $(r,s)=(1,0)$.   The parameters increase iteratively, such that for $l+1$, first we try to increase $s$ by 1, but if $s=r$ we instead increase $r$ by one and reset $s$ to zero.  The relation between the level number and the distance parameters can be determined with simple calculation to be $ l=s(r+1)/2+r $.  The decoder performs the following, beginning with the first level $l=1$:
\begin{enumerate}
\item \textit{Clustering}:  Find a $(r,s)$-partition of $\mathcal{W}_{l}$;
\item \textit{Neutral annihilation}: For every neutrally-charged cluster in the partition, for instance $ \mathcal{C}_{j}$, find a Pauli correction $e_{j}$ with support entirely on $\mathcal{C}_j$ that annihilates all the syndromes within the cluster. 
\item \textit{Refresh:}  Record the collective Pauli correction $ \prod_{j} e_{j}$ and update the syndrome information to $\mathcal{W}_{l+1}$.  If $\mathcal{W}_{l+1}$ is non-empty, then repeat at next level $l=l+1$.
\end{enumerate}

It is helpful to refer to individual levels of the decoder as sub-protocols that we label $\mathfrak{D}_{l}$.  Any charged cluster that cannot be annihilated completely by $\mathfrak{D}_{l}$,  is therefore left for the next higher level of decoding $\mathfrak{D}_{l+1}$. The higher levels will have larger regions and therefore any charged clusters will eventually be combined and form bigger neutral clusters which can then be annihilated. Also, notice that in the HDRG construction the correction chains are determined during the neutral annihilation step at every level of decoding. In classical coding theory, this is a typical feature of what is known as a \textit{hard-decision decoder} \cite{M06,P00}. Later, in Sec. \ref{SecEnhance}, we will introduction an initialization step that is not part of the above main loop and occurs before $\mathfrak{D}_{1}$. The initialization step performs some syndrome manipulation to edit $\mathcal{W}$ prior to use in the first level of the decoder.

There are several crucial difference between our version of the HDRG decoder described above and the original decoder by Bravyi and Haah \cite{BH11}. First, the distance measure in \cite{BH11} is the Max distance $D^{(\infty)}$. Recall that a ball of radius $r$ in this the Max distance is denoted $B^{\infty}_{r}$, and Bravyi and Haah use such a region at level-$r$ of the decoder where $r$ grows exponentially with the decoder level. The refined metric we choose will include all such regions, since $\mathcal{R}_{r, 0} = B^{\infty}_{r}$, but our protocol will consider additional intermittent levels. We choose also a metric which grows linearly as the decoding level increases. Second, their decoder declares failure and aborts if the area of a cluster is larger than half the lattice size. The idea behind this requirement is that annihilating such large clusters would very likely lead to a logical error.  However, in our decoder we did not enforce this requirement because, as we will see, in higher dimensions the syndrome tend to percolate at high enough error probability, and we would like to investigate how this decoder behaves in such regimes. For this reason, in our implementation here, in the annihilation step  we simply combine the syndromes within a cluster with their first possible local neighbors.  Similar variations of the HDRG decoder has been proposed very recently in \cite{W13,BBD13,WBIL13}. 

The run-time complexity of the HDRG decoder depends on the number of non-trivial syndromes. In higher dimensions the number of syndromes increases with the qudit dimension. To understand this behavior consider the following example. In the qubit case if two neighboring errors occur then the shared syndrome will not be detected. But in the qudit case, the probability of two neighboring errors with opposite weights to occur will diminish quickly as the dimension increases. Hence, the shared syndrome will almost always be detected. The consequence of this observation is that for a given error rate the \textit{density} of the syndromes will approach the density of the errors as the dimension increases. Therefore, the exact run-time complexity needs to capture the relation between the number of syndromes and the qudit dimension, which not a trivial task.
Nevertheless, our numerical analysis shows that this dependence on the qudit dimension is negligible and for practical purposes can be ignored.

\subsection{Threshold Estimation}

\begin{figure}
  \centering
    \includegraphics[width=0.4\textwidth]{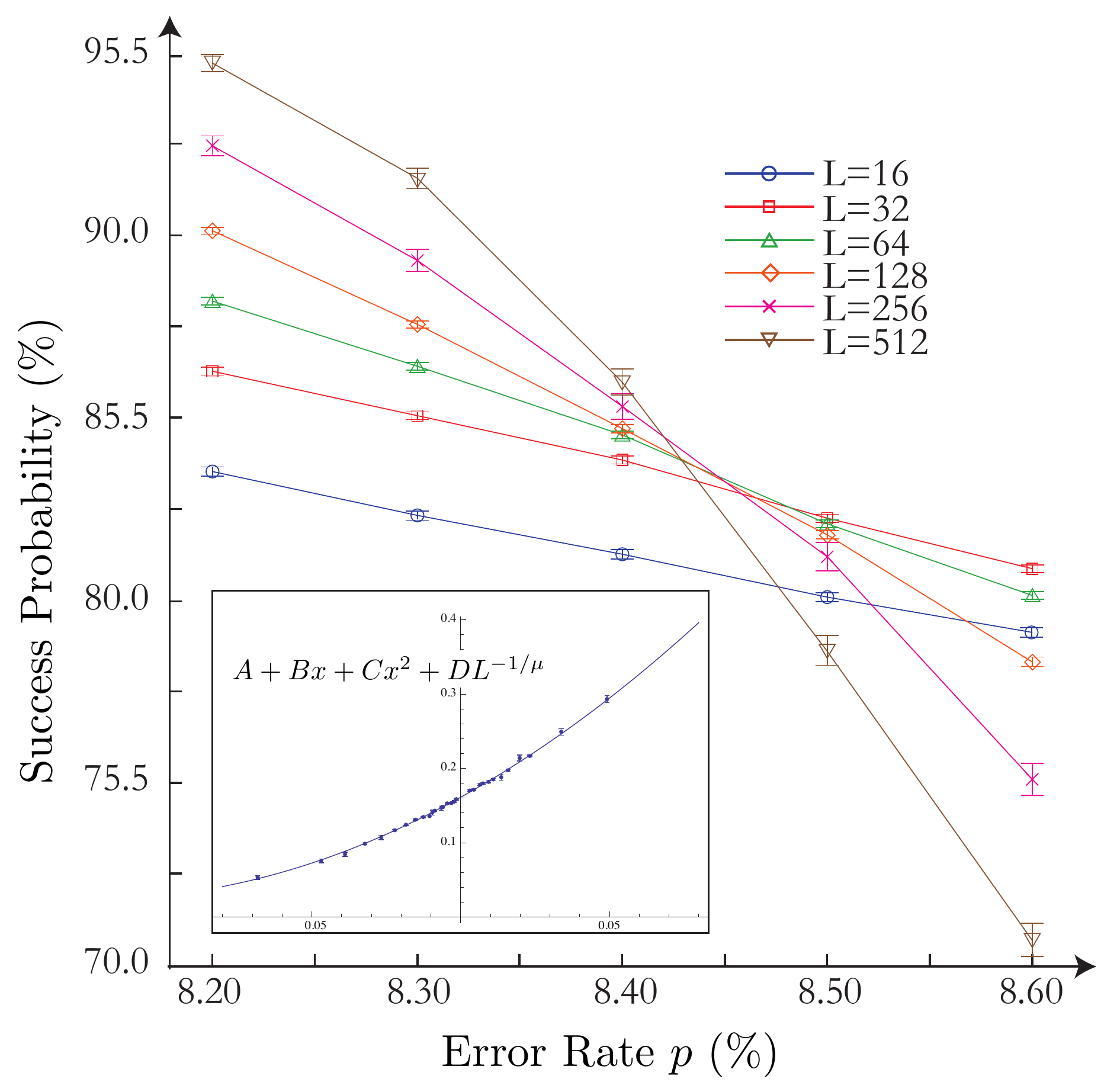}
    \hspace{-5mm}
      \captionsetup{singlelinecheck=yes, justification=centerlast}
      \caption{The success probability of the HDRG decoder for the qubit case. The data points are generated with $ N=10^{5} $ samples for $ L\in\{16,32,64,128\} $ and  $ N=10^{4} $ samples for $L\in\{256,512\} $. The error bars are are taken to be $ 2\sigma $. The boxed plot shows the data fitting, where $ x=(p-p_{th})L^{1/\nu} $, $ \nu=1.85\pm0.04 $ and $ \mu=0.46\pm0.06 $. }
      \label{QubitPlot}
      \vspace{-.5cm}
\end{figure}

In this section we present the results of the Monte Carlo simulation for the HDRG decoder. We begin with the qubit case before moving to higher dimensions. We plot the success probability curves for the qubit case in Fig.~\ref{QubitPlot}. Using the fitting described in Eq. (\ref{HarringtonFit}), we estimate the threshold to be $ 8.4\%\pm0.01 $.

\begin{figure}
  \centering
    \includegraphics[width=0.5\textwidth]{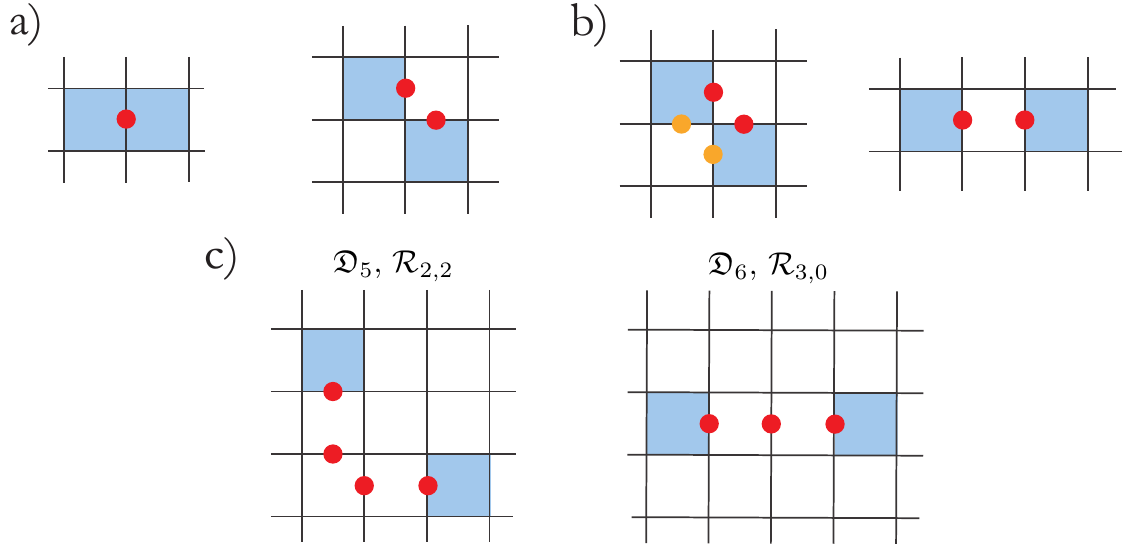}
      \captionsetup{singlelinecheck=yes, justification=centerlast}
      \caption{The regions $ \mathcal{R}_{r,s}$ distinguishes between a) and b), whereas the $D^{\infty} $-metric does not. c) The optimal distance measure will switch levels $ 5 $ and $ 6 $.} 
      \label{Degeneracy}
      \vspace{-.5cm}
\end{figure}

Recall that the threshold achieved by the original HDRG decoder in~\cite{BH11} was $ 6.7\% $. The improvement in the threshold achieved by our HDRG decoder is mainly due the refined set of regions $ \mathcal{R}_{r,s}$, which we have adopted in favor over using just the Max distance. To demonstrate this point, we consider two simple examples. Fig.~\ref{Degeneracy}(a) shows two plaquettes created by one and two errors. Clearly the single error is more likely to occur in comparison to two neighboring errors. However, with the $ D^{\infty} $-metric the plaquettes in both cases will be connected at the first level. But the regions $ \mathcal{R}_{r,s}$ distinguish between the two cases, and they will be connected at two separate decoding levels, namely $ \mathfrak{D}_{1}$ and $ \mathfrak{D}_{2}$. Also, Fig.~\ref{Degeneracy}(b) shows two cases of two plaquettes created by two errors. For the first case, there are two errors for which the set of successful recovery operations are identical.  Hence, the first case is more probable to occur since it has double the degeneracy. The regions $ \mathcal{R}_{r,s}$ better account for this degeneracy by again treating these cases into two separate levels, namely $\mathfrak{D}_{2}$ and $ \mathfrak{D}_{3}$. The overall effect of such refinement is to create finer clusters which would lead to better error correction during the annihilation step.

The above observations suggest that to improve our decoder further one can consider a different sequence of regions.  Such distance measure is optimal in the sense that it will always connect syndromes that can be created by fewer errors and higher degeneracy first. It is not hard to see that such improvement will switch, for example, level $ \mathfrak{D}_5$ with level $ \mathfrak{D}_6$, because the latter will connect syndromes created by fewer errors as shown in Fig.~\ref{Degeneracy}(c).  Our approach, however, was easier to implement, and we leave such further improvement for future investigation.

\begin{figure}[h!]
  \centering
    \includegraphics[width=0.5\textwidth]{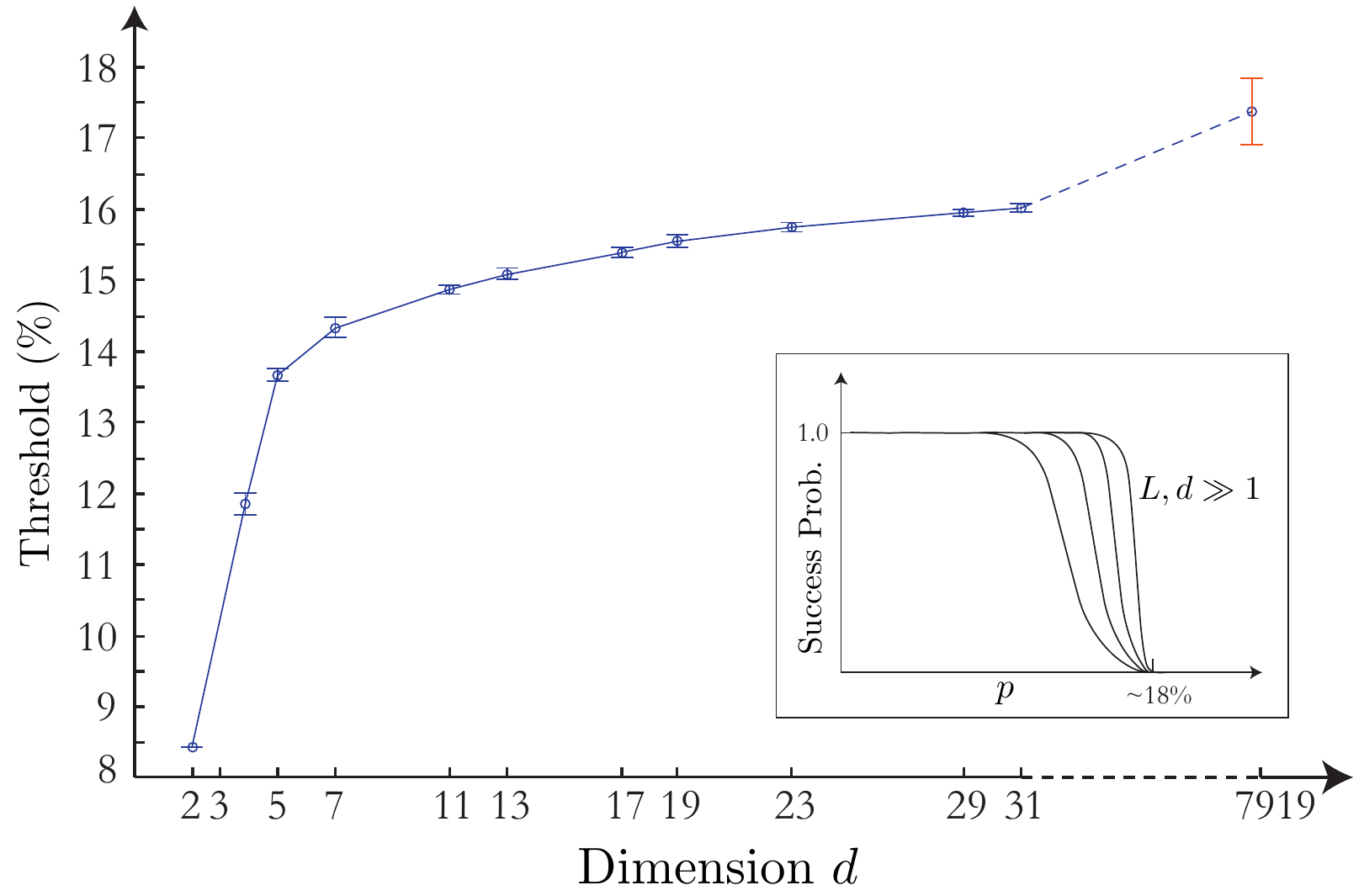}
    \hspace{-5mm}
      \captionsetup{singlelinecheck=yes, justification=centerlast}
      \caption{The threshold values of the HDRG decoder for prime dimensions with $ 2\sigma $ error bars. The boxed plot is illustrative figure of the behavior the success probability in the limit of high $ d $.}
      \label{Alld}
      \vspace{-.5cm}
\end{figure}

The thresholds of the remaining prime dimensions are plotted in Fig.~\ref{Alld}. To demonstrate that a numerical estimate of the threshold can be obtained for any dimension we have chosen the $ 1000 $\textit{th} prime number $ d=7919 $ to represent the limit of high $ d $. As can be seen from Fig.~\ref{Alld}, the thresholds increases monotonically with qudit dimension and reaches a saturating value of about $ 18\% $. We have discovered that the reason for this behavior is due to a percolation effect, which we will discuss next. 

It was pointed out in Sec. \ref{SecToric} that for a given error rate the density of syndromes increases as the dimension of the qudits increase. In fact, as we will show in the next section, for any given prime dimension $ d\ge3 $, there exists a unique threshold error rate at which the syndromes percolates the lattice. In other words, above this threshold the syndromes will always span the lattice completely. We refer to this threshold as the \textit{syndrome percolation threshold}, denoted here by $ p^{\text{syn}}_{\text{th}} $. We will provide numerical estimates of this threshold in the next section. We will find that it decreases as the dimension increases until it reaches a constant value of about $ 18\% $ in the limit of high $ d $, see Fig.~\ref{PercolationGraph}.  

The syndrome percolation has sever consequences for the HDRG decoder. For any error rate $ p>p^{\text{synd}}_{\text{th}} $ there will be one percolating neutral cluster at the first level $ \mathfrak{D}_{1}$ of decoding. The HDRG will try to annihilate the syndromes with their nearest-neighbors and will always fail. This suggests that we cannot expect the HDRG decoder to achieve a threshold higher than the percolation threshold, because the success probability curves must diminish above the percolation threshold. Indeed this is what we observe in the limit of high $ d $, as illustrated by the boxed plot in Fig.~\ref{Alld}. The point of intersection of the curves (which defines the threshold) intersects $ x $-axis at the value of the percolation threshold. The actual curves (omitted here) are too noisy around the syndrome percolation threshold, for this reason we have indicated by the red error bar the range at which the actual curves intersects in the limit of high $ d $. Furthermore, our numerical analysis show that if we ignore the small lattice sizes, then the curves of the large lattice sizes clearly cross at single point around $ 18\%$.

The conclusion of the above discussion is that in the limit of high $ d $ we expect the syndrome percolation threshold to be a close upper-bound to the threshold achieved by the HDRG decoder. Next, we outline some basic concepts in percolation theory and show how the syndrome percolation thresholds can be estimated. 

\subsection{Syndrome Percolation Thresholds}\label{SecSyndPer}

Percolation theory is the study of connectivity and transport on random graphs \cite{BH57,G89,SA94}. A standard percolation model consists of a random graph whose {\em sites} are distributed in space, and the {\em bonds} connect neighboring sites only. We are mainly interested in the percolation behavior on a 2D regular graph, and in particular the regular square graph. There are typically two stochastic mechanisms associated with each graph structure: either the sites of the graph are fixed in space and bonds are made randomly on them, or sites are random in space and the bonds are determined on the neighboring sites.  

For instance, in the random vertex model,  each site is ``empty" with probability $p$  and otherwise it is ``occupied".  For each instance, percolation occurs if there is a nearest neighbor path that spans the graph using only occupied sites.  The key result of percolation theory is that there exists a threshold, $p_{\text{th}}^{\text{site}}=59.27\%$ \cite{G89}, above which the probability of percolation approaches unity with increasing lattice size, and below threshold the percolation probability vanishes in the large lattice limit.  A similar phenomena occurs when randomly removing graph bonds, and has an analytic threshold of $p_{\text{th}}^{\text{bond}}=50\% $. 

On the square lattice of the toric code, the bonds correspond to the qudits on the edges of the lattice and the sites correspond to the vertices/plaquettes operators. In our discussion here, we are interested in the \textit{syndrome} percolation threshold of the toric code. This is  not equivalent to site percolation because the syndromes created in pairs by qudit errors on the lattice edges.  Given a syndrome $\mathcal{W}$ we say that it percolates the lattice if there is a nearest neighbor path in $\mathcal{W}$ that spans the lattice.  Using our terminology, a nearest neighbor path is a $(1,0)$--path in $\mathcal{W}$.  There have been studies of site percolation with distant neighboring interaction \cite{MG05,MM07}, but to our knowledge there have not been investigations where bonds interact with sites in the manner defined by the toric code. Also, there does not appear to be an analytic method that can determine the syndrome percolation threshold precisely from the known theory on the bond and site percolation. 

We resort to estimating the syndrome percolation numerically via Monte Carlo simulations. The simulation is straight forward and it is very similar to that described in Sec. \ref{SecNoise} in estimating the error correction threshold of a general decoder. For a given dimension $ d $, error rate $ p $, and lattice size $ L $, we generate a qudit lattice such that each qudit suffers an error with probability $ p $. The syndromes are then calculated. If the syndromes percolate, then the simulation will be declared successful, otherwise it is a failure. This procedure is then repeated $ N $ times, and the success probability is evaluated as the fraction of times the simulation has succeeded. The simulation is repeated for fixed range of $ p $ for different lattice sizes. The threshold is determined as the point of intersection of the different success probability curves.

\begin{figure}
  \centering
    \includegraphics[width=0.5\textwidth]{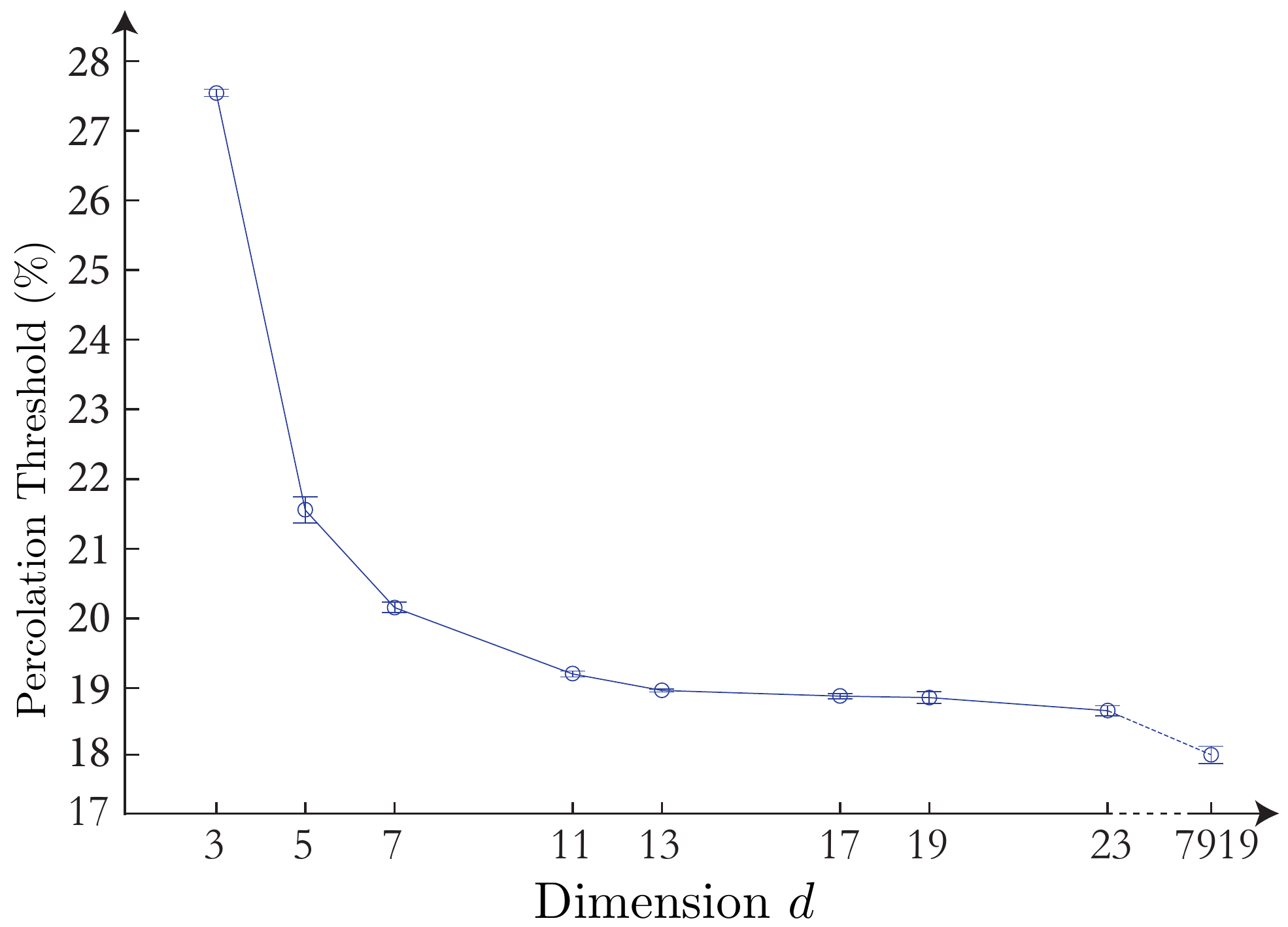}
    \hspace{-5mm}
      \captionsetup{singlelinecheck=yes, justification=centerlast}
      \caption{Syndrome percolation threshold for prime dimensions with $ 2\sigma $ error bars.}
      \label{PercolationGraph}
\end{figure}

The numerical estimates for syndrome percolation thresholds obtained are presented in Fig.~\ref{PercolationGraph}. As can be seen from this figure, there does not exist a syndrome percolation threshold for the qubit case, a fact that can be understood as follows. Consider the probability that a plaquette (the toric code analog of a site) is non-trivial given that the four neighboring qubits independently suffer an error with a probability $ p $. This probability can easily be shown to be $ 4p(1-p)^3 + 4p^3(1-p) $. This expression is symmetric about $ p=0.5\pm c $, , where $ c $ is a constant between $ 0\le c\le 0.5 $. This indicates that the profile of the success probability curve versus the error rate $ p $ has a bell shape about $ p=0.5 $, and therefore prohibiting the existence of a unique threshold point above which the lattice always percolates. However, for the remaining prime dimensions, such symmetry does not exist and we always observe a threshold. We see that the syndrome percolation threshold decreases monotonically with the qudit dimension, and in the limit of high $ d $ it reaches a constant value of about $ 18\% $. This confirms the conclusion of the last section in that the syndrome percolation threshold is an upper-bound for the HDRG decoder. In the next section we will show how the HDRG decoder can be enhanced to beat this upper-bound. 

\subsection{Beating the Percolation Threshold}\label{SecEnhance}

To overcome the syndrome percolation threshold we introduce an \textit{initialization step} $ \mathcal{I}_{r,s} $ that enhances the performance of the HDRG decoder. Although this step is not efficient, we show that for a sufficiently high $ d $ it can boost the threshold to about $ 30\% $ at a computationally feasible cost. The initialization step is designed to dissect any percolating cluster into a more sparse set of clusters \textit{before} running the HDRG decoder. It achieves this by using a brute force method in finding any neutral \textit{sub-clusters} within a percolating cluster. The sub-clusters are then annihilated before running the HDRG decoder. We have constructed the initialization step to search for the sub-clusters systematically by utilizing similar concepts as those used in the HDRG decoder. For example, the subscripts  $ r $ and $ s $ of each step $ \mathcal{I}_{r,s} $ takes the same increasing  integers as was previously defined, and here they quantifies the \textit{depth} of searching for the neutral sub-clusters in the lattice.    

\begin{figure}
  \centering
    \includegraphics[width=0.4\textwidth]{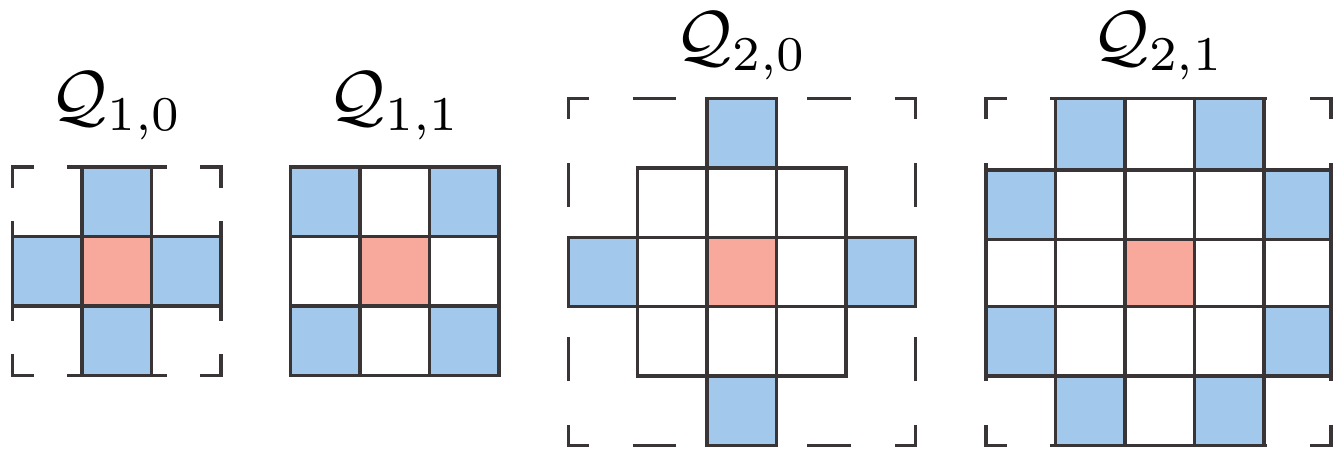}
      \captionsetup{singlelinecheck=yes, justification=centerlast}
      \caption{The set of syndromes $ \mathcal{Q}_{r,s} $ for the first four initialization levels $ \mathcal{L}_{r,s} $. The red square is the syndrome $ u $ and the blue squares are the $ q_{k} $ syndromes at the outer layer of $ \mathcal{R}_{r,s}$.}
      \label{Initialization}
      \vspace{-.5cm}
\end{figure}

Each initialization step $ \mathcal{I}_{r,s} $ consist of a series of \textit{initialization levels} $ \mathcal{L}_{r,s} $ that systematically search for neutral sub-clusters. More precisely, in this construction, each step $ \mathcal{I}_{r,s} $ simply involves running all the initialization levels in ascending order such that $ \mathcal{I}_{r,s} = \{\mathcal{L}_{1,0},\mathcal{L}_{1,1},\dots,\mathcal{L}_{r,s}\}$. Loosely speaking, the subscripts $ r $ and $ s $ of each level $ \mathcal{L}_{r,s} $ quantifies the `size' of the regions to search over for any neutral sub-clusters, we will expand on this point shortly. Each level $ \mathcal{L}_{r,s} $  is associated with a \textit{set} of syndromes $\mathcal{Q}_{r,s} $. The set $\mathcal{Q}_{r,s} $ consists of the syndromes at the outer layer of $\mathcal{R}_{r,s}$, such that
\begin{equation}
	\mathcal{Q}_{r,s}(\vec{x}) = \bigg\{ \begin{array}{l}
	  \mbox{for } s=0, \mathcal{R}_{r,s} \setminus     \mathcal{R}_{r-1,r-1} , \\ 
  	  \mbox{for } s>0, \mathcal{R}_{r,s} \setminus     \mathcal{R}_{r,s-1} ,
	\end{array}
\end{equation}
where "$A  \setminus B$" just means in $A$ but not in $B$.  This is more easily shown by the examples in Fig.~\ref{Initialization}. We denote the elements of the set $\mathcal{Q}_{r,s}$ by $ q_k $, and by definition, each set has either $ 4 $ or $ 8 $ number of syndromes. Next, we describe how the searching procedure at each initialization level works, and again without loss of generality we will limit the discussion to the plaquette operators. We will denote the set of all plaquettes by $ \mathcal{U}=\{u_{1},u_{2},\dots,u_{L^{2}}\} $.

At each level $ \mathcal{L}_{r,s} $, the search for sub-clusters is performed by starting at a plaquette $ u_{j} $ in the lattice (regardless if it is charged or not) and then for each $ q_k\in\mathcal{Q}_{r,s}$, we construct a \textit{search rectangle} $ \mathcal{T} $, which is defined as the \textit{minimum} size rectangle that encloses syndromes $ u_j $ and $ q_k $. In other words, the plaquette $ u_{j} $ and $ q_k$ form the opposite corners of the search rectangle. Inside $ \mathcal{T} $, we define a \textit{search-path} $ \tau $ as any $ (1,0)$--path in $ \mathcal{T} $ that starts at $ u_{j} $ and ends at $ q_{k} $. There are many such paths, and by construction, every path will contain $ |\tau|=(r+s+1) $ elements in total. We denote the set of all possible search-paths in $ \mathcal{T} $ by $ T=\{\tau_1,\dots,\tau_{|T|}\} $, where $ |T| $ is the total number of possible paths. From a pure geometric point of view, a rectangle consisting of $ a\times b $ plaquettes has $|T|=(a+b)!/a!b! $ possible paths connecting its corners. This expression was calculated by considering the equivalent problem of finding all the minimum paths between two points on a Manhattan geometry \cite{G97}. The idea here is to treat each search-path as an independent sub-cluster, and the aim is to annihilate any neutral sub-clusters.

Based on the above definitions, we now summaries the searching routine of an initialization level $ \mathcal{L}_{r,s} $ as follows. For each plaquette $ u_{j}\in\mathcal{U} $ (starting with $ u_{1} $):
\begin{enumerate}

\item Choose an element $ q_j\in\mathcal{Q}_{r,s}$, and construct a search rectangle $ \mathcal{T} $;
\item Search for all possible sub-clusters $ \tau_j\in T $ within $ \mathcal{T} $ systematically. If any sub-cluster $ \tau_j $ is found to be neutral, then annihilate $ \tau_j $ and stop the search. Then start step $ 1 $  with the next plaquette $ u_{j+1}\in\mathcal{U} $; Else
\item If no neutral sub-cluster were found, choose the next element $ q_{j+1}\in\mathcal{Q}_{r,s} $ and repeat steps $ 1 $ and $ 2 $; Else
\item If there are no remaining syndromes $ q_j\in\mathcal{Q}_{r,s} $, then the search has ended without finding a neutral sub-clusters for plaquette $ u_{j} $. Start step $ 1 $  with the next plaquette $ u_{j+1}\in\mathcal{U} $.
\end{enumerate}

The above procedure is repeated until all the plaquettes $ u_{j}\in\mathcal{U} $ have been searched. The overhead of this search procedure is proportional size of the search rectangle $ |T| $, which is factorial in $ r$ and $ s $. More precisely, for each initialization level $ \mathcal{L}_{r,s} $, in the worse case scenario (where no neutral sub-clusters are found) the search takes $ \alpha L^{2} $ steps, where the constant overhead $ \alpha=(r+s)!/r!s! $. Although that seems to be completely inefficient, the parameters $ r $ and $ s $ increase polynomially with the number of initialization levels, and hence for the first few levels $ \alpha $ is small enough. As a result, running the above procedure for the first few levels is still a computationally feasible task. It is important to notice that for each plaquette $ u_{j} $ the procedure stops once a neutral sub-cluster is found, and the worst case of not finding any neutral sub-clusters happens only when the dimensions $ d $ and the error rate $ p $ are sufficiently high. 

The depth of searching for the neutral sub-clusters increases as the initialization levels increase in size. We propose an \textit{enhanced}-HDRG decoder at depth $ (r,s)$ to consists of running the initialization step $ \mathcal{I}_{r,s} $ followed by the HDRG decoder described in Sec. \ref{SecHDRG}. 

\begin{figure}
  \centering
    \includegraphics[width=0.5\textwidth]{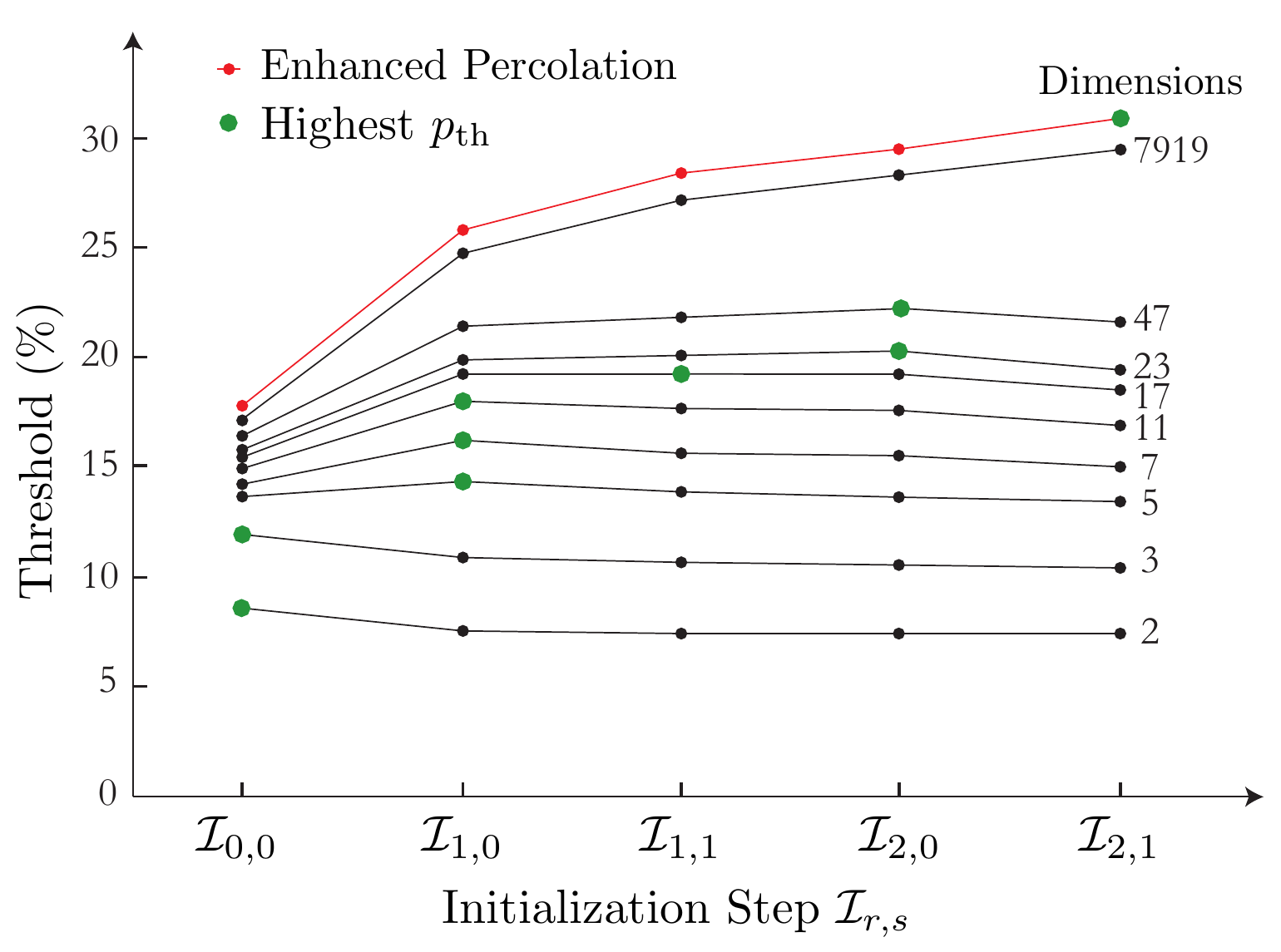}
    %\hspace{-5mm}
      \captionsetup{singlelinecheck=yes, justification=centerlast}
      \caption{The thresholds for the enhanced-HDRG decoder with the first four initialization steps $ \mathcal{I}_{r,s}$. The error bars and data of some prime dimensions are not included for clarity. The red curve is the enhanced-syndrome-percolation threshold in the limit of high $ d $.}

      \label{FinalFig}
\end{figure}

The numerical estimates for the thresholds achieved by the enhanced-HDRG decoder for the first four initialization steps are summarized in Fig.~\ref{FinalFig}. The thresholds for $ \mathcal{I}_{0,0}$ corresponds to the HDRG decoder without any enhancement, with the corresponding thresholds previously presented in Fig.~\ref{Alld}. For the qubit and qutrit cases we see that the thresholds decreases after the initialization steps are introduced. This is because for these low dimensions, finding a neutral sub-cluster is very probable, and hence the initialization step is in fact too destructive. As a result the clusters are divided into very sparse set of smaller clusters, and running the HDRG will end up connecting these sparse sets of clusters and causing more logical errors. 

However, we start to observe improvement in the thresholds above the qutrit case. Notice that for all the listed first few primes dimensions, after some initialization step the thresholds start to decrease. This is also because after some depth of searching the initialization step becomes too destructive. In the limit of high $ d $, we see that a threshold just under $ 30 \% $ can be achieved. The current shape of the curve indicate a potential increase in threshold with initialization step beyond $ \mathcal{I}_{2,1}$, and we leave such investigation for a future work.  

Finally, in the limit of high $ d $, the saturating thresholds of the enhanced-HDRG decoder can also be explained by the syndrome percolation effect. We introduce the {\em enhanced}-syndrome-percolation threshold which is determined by simply running the initialization step $ \mathcal{I}_{r,s}$ followed by the syndrome percolation simulation described in Sec. \ref{SecSyndPer}. The numerical estimates for the enhanced-syndrome-percolation thresholds are presented in Fig.~\ref{FinalFig} by the red curve. Our numerical analysis shows that the enhanced-HDRG decoder can reach the upper-bound of the red line by ignoring small size effects and considering large lattice sizes only. 

\section{{\em soft-decisions}-RG Decoder} \label{SecRG2}

\subsection{SDRG Overview}
In this section we study the SDRG decoder introduced by Duclos-Cianci and Poulin in \cite{CP10,CP102}. The SDRG decoder used here, developed independently of that used by Duclos-Cianci and Poulin in Ref.~\cite{CP13}, differs from their approach in that we optimize the decoder for very high speed decoding at the expense of a reduced threshold. This enables us to probe thresholds up to very large $d$. In this section we broadly review the techniques used in the SDRG decoder. Next, we introduce the specific implementation of the SDRG decoder we use. Finally, we discuss the thresholds obtained by this decoder.

It would be cumbersome to describe this decoder without employing homology terminology. 
In the following, for the non-expert reader, {\em homological equivalence} can be taken as equivalent to equivalence under multiplication by a member of the stabilizer group (or more precisely the stabilizer subgroup generated only by plaquette or vertex operators, depending on context). Two homologically equivalent objects are referred to as being {\em homologous}. A homology class, is an equivalence class of operators equal up to a member of the vertex or plaquette stabilizer subgroup (as appropriate).  We refer to the reader who would like a precise definition of these terms to App.~\ref{HomologyAppendix}. 

\begin{figure}
\includegraphics[width=\columnwidth]{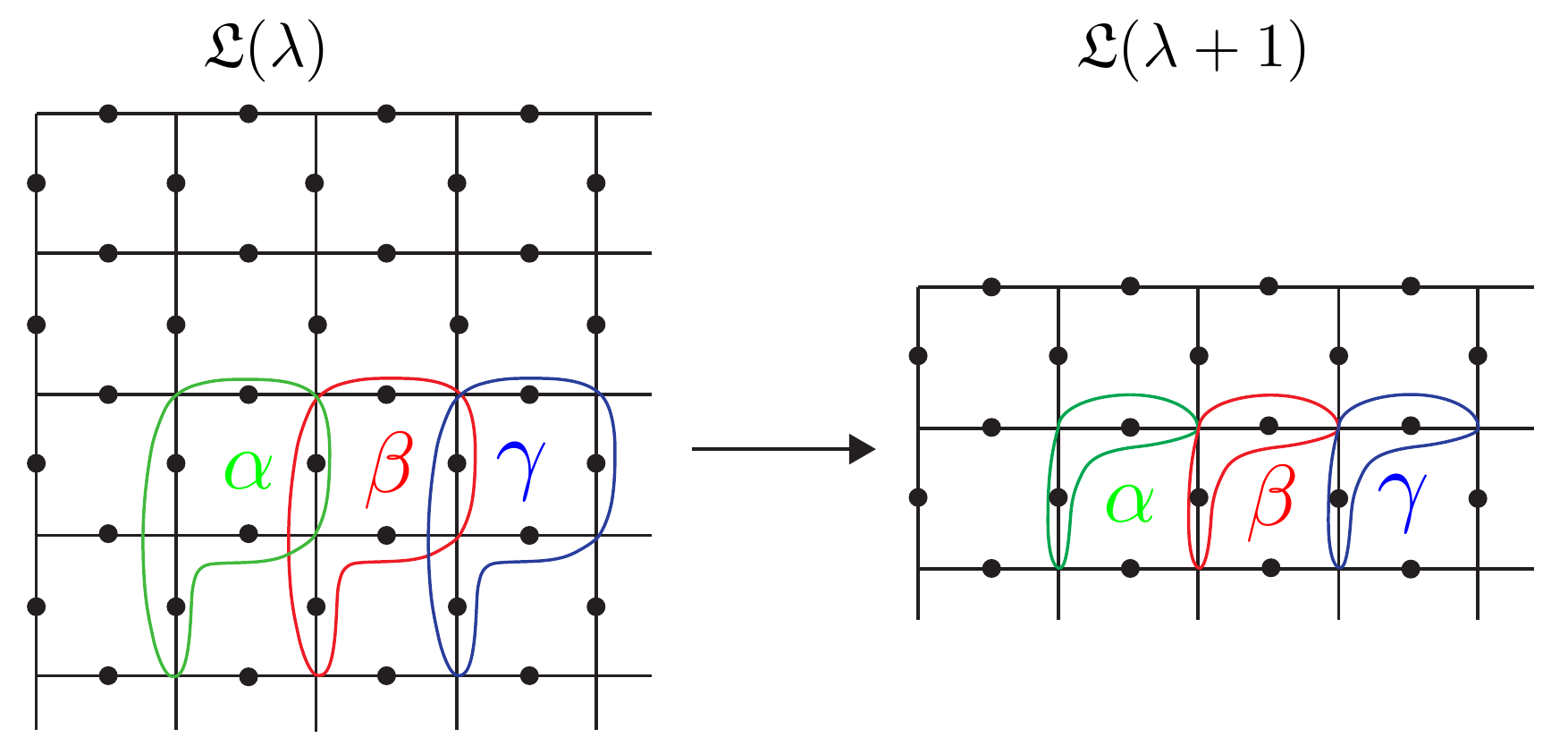}
\caption{\label{RGLattices} Three fixed, overlapping $2\times 1$ cells, $\alpha$, $\beta$ and $\gamma$, of a $4 \times 4$ lattice. The cells coarse grain $\mathfrak{L}(\lambda)$ to the $4 \times 2$ lattice $\mathfrak{L}(\lambda +1)$ in the implementation used here.}
\end{figure}

The SDRG decoder has a run-time complexity $O(L^2 \log L)$. It works by approximating the relative likelihood of different {\em homology classes} $\mathcal{H}$ of error configurations $e$ with corresponding error operators $E= X(e)$, where we are using the notation
\begin{equation}
X(e) = \bigotimes_{j} X_{j}^{e_{j}} ,
\end{equation}
where $e=\{ 0,\dots,d-1 \}^{n}$ is an $ n-$dimensional vector.

We evaluate the probability of a homology class of error configurations $\probfont{P}_\mathcal{H}$ by summing over probabilities of error configurations $\probfont{P}(e)$ for all $e \in \mathcal{H}$ 
\begin{equation}
    \probfont{P}_{\mathcal{H}} = \sum_{u \in \mathcal{H}} \probfont{P}    ( u ).
\end{equation}

On the torus, we have $d^{2}$ distinct homology classes. Homology classes differ by addition of configurations of non-contractible loops, $l$, where, for example, we may have $l$ such that $\bar{X}_{1}=X(l)$. The calculated probabilities of all homologous $u$ for all $\mathcal{H}$ can then be used to produce a correction operator from the appropriate homology class to attempt to return the lattice to its initial state. This method of exhaustive decoding is not adopted because it is not efficient with system size. We find all the elements of a homology class by stabilizer deformations on the qudit toric code, where we have $O(L^2)$ stabilizers, we have therefore $O(d^{L^2})$ elements of every homology class.  Summing over an exponential number of correction operators is clearly inefficient.

It is not necessary to consider all the error configurations within a homology class. Instead, we can consider probabilities of many `sensible' error configurations which are likely to have occurred and still achieve respectable thresholds. The SDRG decoder uses renormalization group methods to efficiently consider the probabilities of many sensible error configurations. It coarse grains syndromes and prior error probability distributions, or {\em priors}, over multiple scales using Bayesian inference methods. We label different length scales with an integer $\lambda$. The decoder coarse grains over  $\sim O(\log L)$ levels, until it reaches the final coarse graining level which we label $\lambda_0$. The priors at level $\lambda_0$ correspond to approximate probabilities of the error configuration on the original lattice having come from particular homology classes.

We denote a lattice which contains both syndromes and priors at different scales by $\mathfrak{L}(\lambda)$. To efficiently coarse grain $\mathfrak{L}(\lambda)$, the SDRG dissects $\mathfrak{L}(\lambda)$ into small fixed {\em cells} of constant size. Each cell occupies a local connected area of $\mathfrak{L}(\lambda)$. Examples of three cells, $\alpha$, $\beta$ and $\gamma$ are shown in green, red and blue respectively in Fig.~\ref{RGLattices}. This cellular decomposition is then used to coarse grain $\mathfrak{L}(\lambda)$ to $\mathfrak{L}(\lambda +1)$, shown on the right of Fig.~\ref{RGLattices}.  Syndromes of the coarse grained lattice $\mathcal{L}(\lambda + 1)$ are evaluated by summing the syndromes of each cell, and the priors of $\mathfrak{L}(\lambda+1)$ correspond to probabilities that the syndrome of the cell is generated by an error chain from different homology classes of the cell. Each cell is decoded exhaustively. As the size of each cell is constant, and small, the time to decode a single cell is constant, and fast. The cells of $\mathfrak{L}(\lambda)$ are decoded in $O(L^2)$ time, with the capacity to be parallelized to constant time. After coarse graining to scale $\lambda_0 \sim O(\log L)$, we arrive at $\mathcal{L}(\lambda_0)$ whose syndrome is necessarily vacuum and whose edge priors contain the probabilities that the syndromes were generated by an error configuration from different homology classes. 

Coarse graining $\mathfrak{L}(\lambda)$ by exhaustively decoding individual cells will only give approximate priors for $\mathfrak{L}(\lambda+1)$, as each cell only has access to restricted local information from the local region of the cell of $\mathfrak{L}(\lambda)$. In particular, at the boundaries where cells dissect the lattice, the approximation used is very poor. To overcome this, the SDRG decoder employs {\em belief propagation} to share information between neighboring cells before renormalization takes place. The cells are chosen such that they contain overlapping edges with neighboring cells, as in Fig.~\ref{RGLattices}. Before the cells are renormalized, they pass {\em marginal messages} to other neighboring cells. The messages take the form of a probability distribution, and describe the beliefs of a cell of what physical errors may have occurred on edges shared, given its syndrome information. In a similar spirit to exhaustively decoding each small cell, the marginal messages are also evaluated exhaustively over the cell in a constant time. Messages received from nearby cells are used to find better priors when the cells are coarse grained. In general, many messages can be shared between cells, where new messages are generated iteratively using previous messages. Multiple iterations of this step significantly enhance the performance of the SDRG decoder.

In the following sections, we describe how renormalization step of the decoder works using messages that we assume have already been exchanged. We then explain how the messages are generated and passed, and we finally discuss the performance of this decoder.

\subsection{Decoder Implementation}

\begin{figure}
\includegraphics[width=0.3\textwidth]{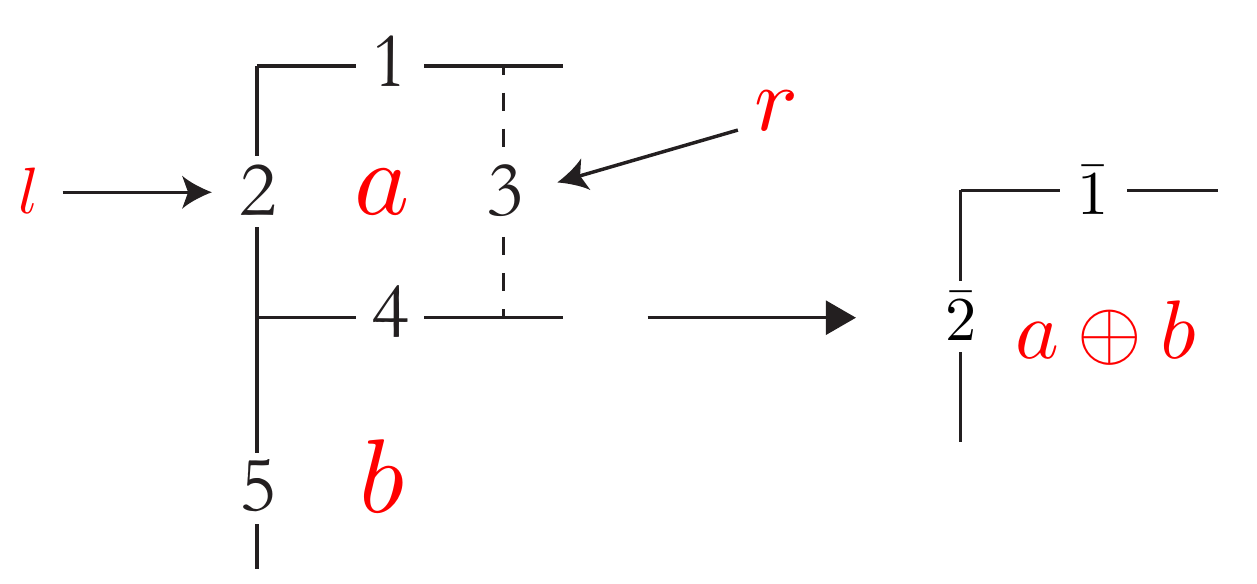}
\caption{\label{RGCell}A five edge cell of $\mathfrak{L}(\lambda) $ which is renormalized to two edges of $\mathfrak{L}(\lambda+1)$. The cell receives messages from its left and right neighbors at $l$ and $r$.}
\end{figure}

The decoder will coarse grain the lattice $\mathfrak{L}(\lambda)$, to a lattice of fewer edges $\mathfrak{L}(\lambda +1)$. The decoder implementation used here, even and odd values of $\lambda$ employ different shape renormalization cells.  For even $\lambda$ we use cells of  $2\times 1$ vertices, and for odd $\lambda$ we use cells of $1 \times 2$ vertices. We describe in detail the coarse graining and belief propagation stages for a $2 \times 1$ cell as shown in Fig.~\ref{RGCell}, but the cell decomposition for odd or even $\lambda$ are equivalent up to a reflection. We note that the cells used here are the smallest possible cells that can be used in such a decoder, which optimize the speed of the algorithm. In choosing this cell size, it is necessary to use different cell shapes at odd and even $\lambda$. A further detail of the message passing stage in the implementation used here, cells pass messages only to left and right neighboring cells for even $\lambda$, and to above and below neighboring cells for odd $\lambda$. In a general implementation however, messages can be passed in all directions at all levels. Cells evaluate probabilities of their own homology classes, which become priors on the coarse grained lattice. The decoder uses many cells at every $\lambda$. However, the action of a single cell of each  $\mathfrak{L}(\lambda)$ is identical up to its input.   In the following subsection we describe in detail the action of a single cell, and its two nearest neighbors, which is repeated over the entire lattice $\mathfrak{L}(\lambda) $ for all $\lambda$. 

\subsection{Renormalization Cells}

Each cell contains five edges, and two syndrome measurements, $a$ and $ b$, which are shown at the left of Fig.~\ref{RGCell}. As before, we denote operators of Pauli $X$ operators with notation $X(e)$ where now error configuration $e$ now only covers 5 edges indexed on the cell shown on the left of Fig.~\ref{RGCell}.

A cell will coarse grain its syndromes. For the plaquette operators we perform this coarse graining by moving syndrome $a$ shown in Fig.~\ref{RGCell} onto the face of syndrome $b$, such that the coarse grained syndrome will take the value $a \oplus b$.  Coarse graining is achieved using the operator
\begin{equation}
  T^a=X(at)=X(0,0,0,a,0).
\end{equation} 
The configuration  $at$ is a member of a homologous class of configurations which will have no errors on the edges of its corresponding coarse-grained cell. We change the class of the coarse-graining configurations to consider the probabilities of errors suffered on the coarse grained edges of a cell using the logical operators
\begin{eqnarray}
\overline{X}_1 &=& X(l_1) = X(1,0,0,1,0), \\
\overline{X}_2 &=& X(l_2) = X(0,0,0,0,1).
\end{eqnarray} 
These configurations modify the class of a coarse-graining configurations because they represent error configurations that extend between different cells.

So far, we have specified three $X(e)$ operators (with $e=t,\,l_1,\,l_2$) in a renormalization cell, but need another two independent operators to form a complete basis for all possible errors.  The remaining two operators are vertex operators truncated to the support of the cell. The are sometime called gauge stabilizers in the literature. They are
\begin{eqnarray}
    S_1 &=& X(s_1) = X(0,1,0,1,d-1), \\
    S_2 &=& X(s_2) = X(0,0,1,d-1,0).
\end{eqnarray} 
Two error configurations $e$ and $e'$ are now homologous if they differ by a gauge stabilizer. Specifically, this is true if and only if there exists $\mu, \nu \in \mathbb{Z}_{d} $ such that $e = e' \oplus \mu s_1 \oplus \nu s_2$.  

The cell only considers error configurations consistent with the syndrome. Given a particular measurement syndrome $a$ we are interested in classes of homologous errors $\mathcal{H}(h_{1}, h_{2})$ for $h_{1}, h_{2} \in \mathbb{Z}_{d}$ where $\mathcal{H}(h_1,h_2) $ is the class containing elements homologous to $t^a\oplus  h_{1}l_1\oplus h_{2} l_2 $. We shall calculate the relative likelihood of each of these classes as described in the next section.

Let us reflect for a moment on the last layer of renormalization.  For $\mathfrak{L}(\lambda_0)$ we have a lattice with only 2 edges and a single syndrome.   The syndrome has been generated by summing syndromes at different levels of renormalization in such a way that it equals the sum of syndromes over the whole lattice.  Since the whole lattice is charge neutral we know that the last syndrome must be trivial, and so syndromes play no further role at this stage.  Whereas the edges and the probabilities of them carrying an error can be directly interpreted as the relative probabilities of each homology class of the original lattice at the microscopic scale, and hence we choose our recovery error from the most likely error class.

\subsection{Coarse Graining Priors}

In addition to syndrome information, each cell contains a set of prior probability distributions and messages received from cells to the left and right.  Each edge, $j$, of $\mathfrak{L}(\lambda)$ contains a prior probability distribution $\probfont{p}_j$ that takes as input $e_{j}\in \mathbb{Z}_d$ and outputs a estimated probability $\probfont{p}_j(e_{j})$ for an $X_{j}^{e_{j}}$ error.  The initial lattice $\mathcal{L}( 0)$ contains the original lattice syndrome and takes its priors from the error model described in Section~\ref{SecNoise}. Each message, $\probfont{q}_{l,r}$, encodes beliefs calculated by neighboring cells which share edges 2 and 3.

Based on these priors, a cell will evaluate $\probfont{p}_{1'}$ and $\probfont{p}_{2'}$ which are coarse grained priors to be used in $\mathfrak{L}(\lambda+1)$.  This is achieved by considering the probabilities of error configurations for different homology classes of the cell.   First we find the probability of a particular error configuration
\begin{equation} 
\probfont{P} (e) = \probfont{p}_1(e_1)\probfont{q}_l(e_2) \probfont{q}_r(e_3)\probfont{p}_4(e_4)\probfont{p}_5(e_5), \label{ErrorConfigurationProbability}
\end{equation} 
which evaluates the probability that an error configuration has occurred using priors $\probfont{p}_j$ and messages $\probfont{q}_{l,r}$.

However, we are actually interested in probabilities over a whole homology class $\mathcal{H}(h_{1}, h_{2})$ and so
\begin{equation} 
\probfont{P}_{\mathcal{H}(h_{1}, h_{2})}= \sum_{u \in \mathcal{H}(h_{1},h_2)} \probfont{P}(u), \label{ErrorConfigurationProbabilitytwo}
\end{equation} 
Ideally, we would pass on all of this information to the next level of renormalization as it represents our belief of the joint probability distribution $\probfont{p}_{1'} \times \probfont{p}_{2'}$.  However, our renormalization cells only accept input priors for individual edges, and these are given by the marginal distributions:
\begin{eqnarray}
    \probfont{p}_{1'}(e_{1}) &=& \sum_{k \in \mathbb{Z}_{d}} \probfont{P}_{\mathcal{H}(e_{1}, k)} , \\       
    \probfont{p}_{2'}(e_{2}) &=& \sum_{k \in \mathbb{Z}_{d}} \probfont{P}_{\mathcal{H}(k, e_{2})} .
\end{eqnarray}
A smarter use of correlations, which we have discarded here, could lead to improved thresholds \cite{privatecomm}.

\subsection{Belief Propagation}

To enhance the performance of the decoder, each cell is supplied with marginal messages from neighboring cells. The messages correspond to the beliefs of a cell that {\em physical} errors have occurred on particular edges. The messages are calculated before each level of coarse graining. We label one cell $\beta$, and its left and right neighbors are labeled $\alpha$ and $\gamma$, as shown in Fig.~\ref{RGLattices}. Each $\beta$ prepares two messages which are the believed error distributions over the shared edges, 2 and 3 of Fig.~\ref{RGCell}, between neighboring cells $\alpha$ and $\gamma$ using the syndrome information of the cell. One message $L$ is passed left to become $\probfont{q}_r$ of cell $\alpha$ and the other, $ R$, is passed right to become its $\probfont{q}_l$ of cell $\gamma$.   Keep in mind that each message is a list of $d$ numbers, e.g. $L$ is communicated as the vector $\{ L(0), L(1),...,L(d-1) \}$.  At the same time $\alpha$ and $\gamma$ will respectively prepare messages $\probfont{q}_l$ and $\probfont{q}_r$ respectively for cell $\beta$. These messages are then exchanged for later message passing rounds or for use in coarse graining. 

At the beginning of any level $\lambda$, all the messages $\probfont{q}_{l,r}$ are initialized to the uniform distribution. The messages are then evaluated to be the marginals over all homology classes
\begin{eqnarray}
  L(e_{2}) &=& \sum_{u \in \mathcal{H}} G(u) \delta_{u_{2}, e_{2}} , \\
  R(e_{3}) &=& \sum_{u \in \mathcal{H}} H(u) \delta_{u_{3}, e_{3}} ,  
\end{eqnarray}
which are in terms functions $G$ and $H$ which return probabilities of error configurations that we define shortly.  Here, $\mathcal{H}$ is the union over all $\mathcal{H}(h_{1}, h_{2})$ and $\delta_{u_{j}, e_{j}}$ is an indicator function that equals unity when $u_{j}=e_{j}$ and zero otherwise. The presence of the indicator function ensures that we are calculating marginal probabilities.   We can unpack this notation, by considering when the indicator function doesn't vanish to find the more explicit, but less compact, formulae
\begin{eqnarray*}
  L(e_{2}) &=& \sum_{h_{1}, h_{2}, \nu } G( h_{1}l_1 \oplus h_{2}l_2 \oplus e_{2} s_1 \oplus \nu s_2 \oplus a t   ), \\
  R(e_{3}) &=& \sum_{h_{1}, h_{2}, \mu } H( h_{1}l_1 \oplus h_{2}l_2 \oplus \mu s_1 \oplus e_{3} s_2 \oplus a t   ) . 
\end{eqnarray*}
Conveniently, the only error configurations we consider that act on edges $2$ and $3$ are $s_1$ and $s_2$, respectively. This is why we are able to change $e_{2}$\,($e_3$) simply by adding the error configuration $s_1$\,($s_2$).  Now, we reveal the functions upon which these equations depend
\begin{eqnarray}
 G(e) &=& \probfont{p}_1(e_1) \probfont{p}_2(e_{2}) \probfont{q}_r(e_{3})  \probfont{p}_4(e_{4}) \probfont{p}_5(e_{5}), \\
 H(e) &=& \probfont{p}_1(e_1) \probfont{q}_l(e_{2}) \probfont{p}_3(e_3)  \probfont{p}_4(e_{4}) \probfont{p}_5(e_{5}). 
\end{eqnarray} 
One may notice from $G$ and $H$ that the messages being passed left (right) do not evaluate new messages using messages received from the left (right). This is to avoid feedback, where messages are created using messages that have previously been sent. 

It is easily seen that the computational complexity of evaluating a round of messages is the same as performing one coarse graining step. However, the improvement in threshold by applying belief propagation significantly enhances the threshold of the decoder, so it pays to spend a few rounds evaluating messages \cite{CP10}. Further, short cuts can be found to evaluate future messages, after the first round of messages have been evaluated by simply performing updates on the previous messages, rather than evaluating new messages from scratch. This significantly speeds up evaluation of messages. In the implementation of the SDRG decoder used here we use five rounds of message passing at each stage before performing a renormalization step. After a few rounds of message passing, messages tend to converge, and they are used to coarse grain the lattice in a renormalization step.

\subsection{Threshold Estimation and the Hashing Bound}\label{SecSDRGHash}

\begin{figure}
  \centering
    \includegraphics[width=0.5\textwidth]{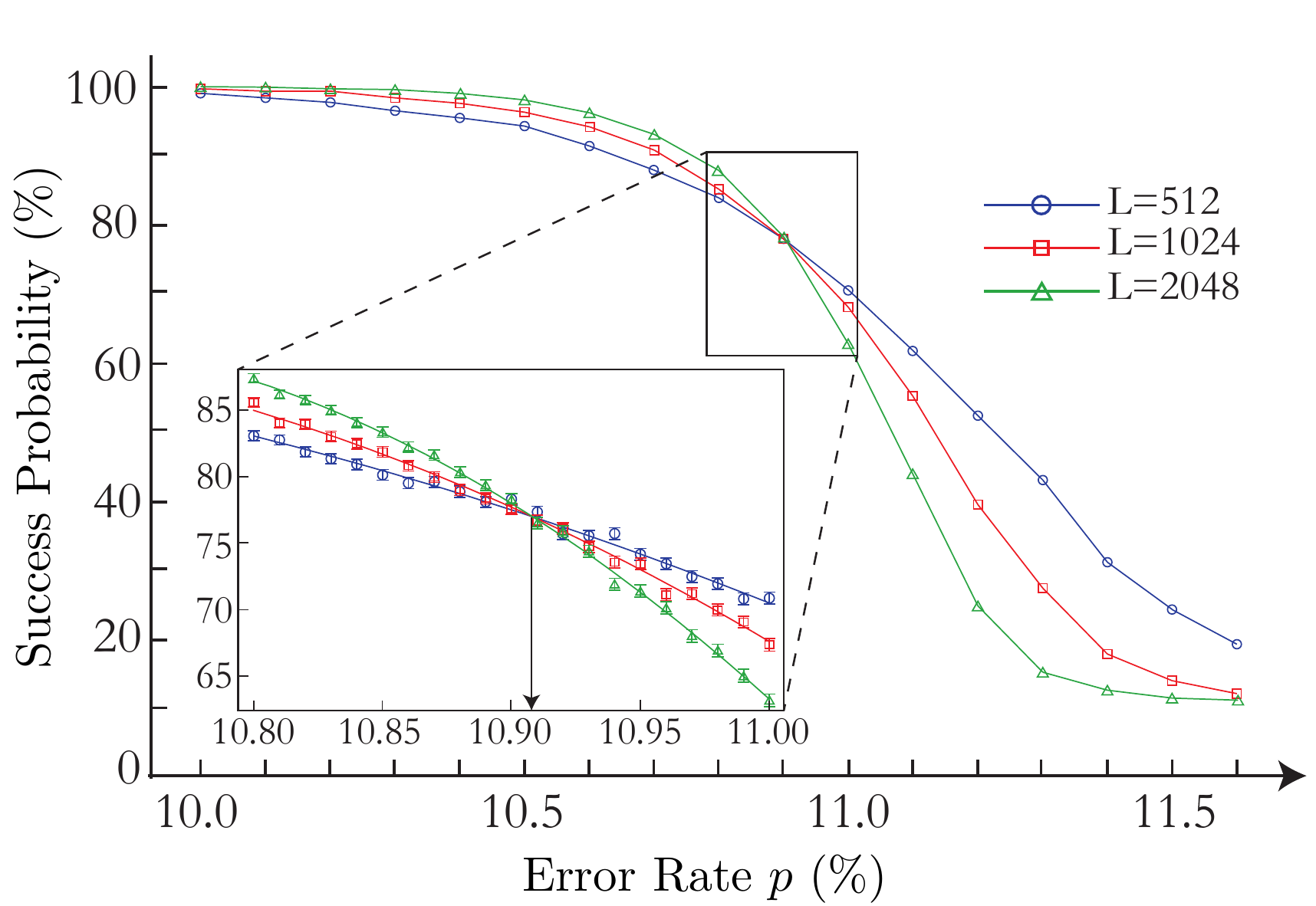}
    \hspace{-5mm}
      \captionsetup{singlelinecheck=yes, justification=centerlast}
      \caption{\label{QutritPlot} Threshold crossing of the SDRG decoder by plotting $p_{s}$ vs. $p$. The inset shows the data points collected close to the crossing point which were fitted to the fitting (\ref{HarringtonFit}). Each data point is calculated using $N = 10^4$ Monte Carlo samples.}   
\end{figure}

\begin{figure}
\includegraphics[width=\columnwidth]{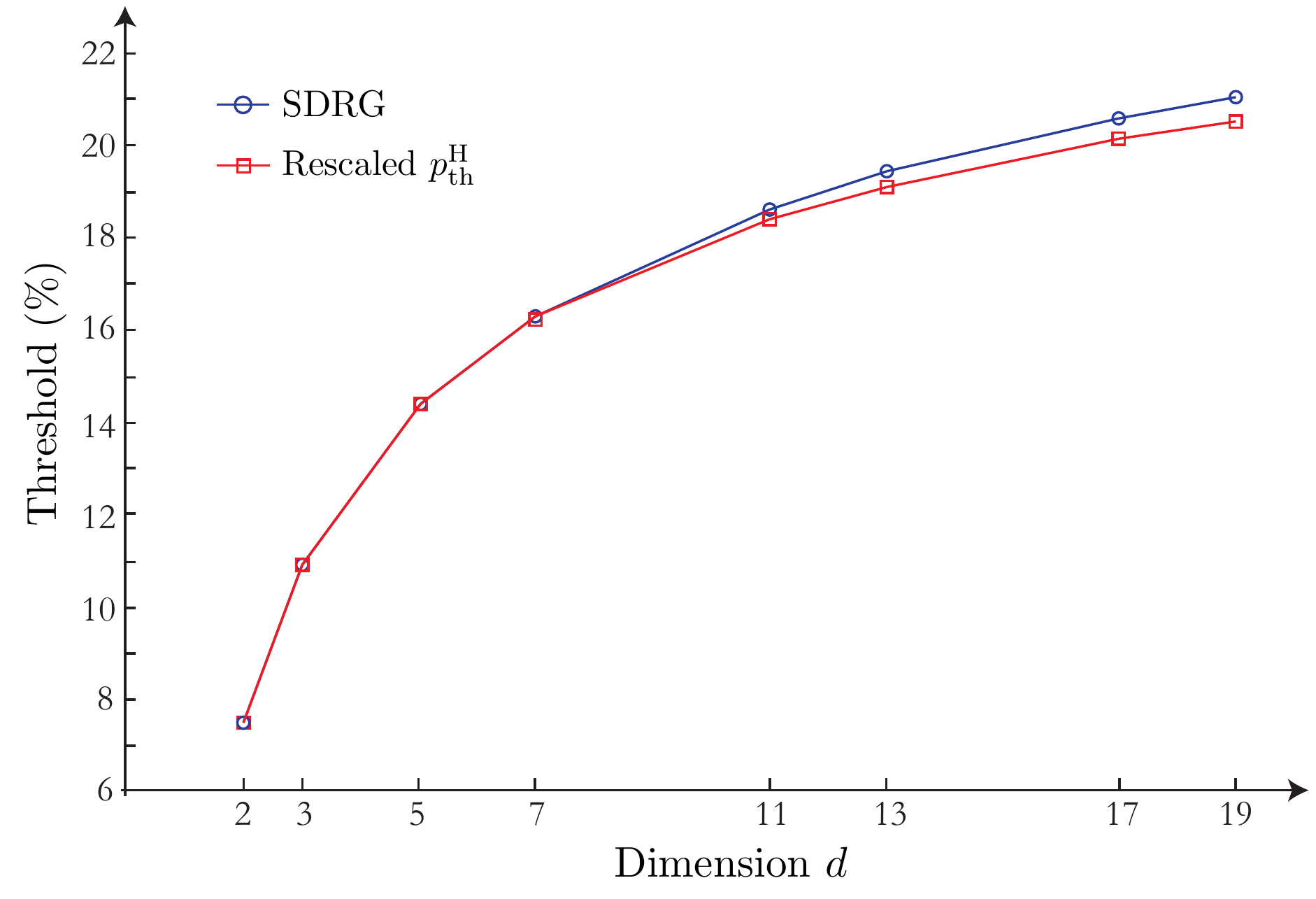}
\caption{\label{Thresholds} Shows thresholds for prime $d$ for the SDRG (in blue), and a constant factor of $\sim 0.68$ of the generalized Hashing bound (in red).}
\end{figure}

In estimating the thresholds we have only considered the crossing of the three largest system sizes to reduce small system size effects, and we used the fitting in Eq.~(\ref{HarringtonFit}). An example is shown for the qutrit case in Fig.~\ref{QutritPlot}, where we use system sizes $L = 512$, $ 1024 $ and $ 2048 $ to find the crossing. The inset of Fig.~\ref{QutritPlot} shows the points close to the crossing point we fit (\ref{HarringtonFit}) to, as well as the fitting itself. We evaluate each $p_\text{succ}$ using $N = 10^4$ samples. We calculate thresholds up to $d = 19$, however, due to the run time complexity the decoder shows in $d$, we reduce system sizes as we increase $d$. The $d = 19$ data point uses system sizes $L = 16$, $32$ and $ 64 $. The achieved thresholds are shown in Fig.~\ref{Thresholds}.

The choice of small cells in the SDRG decoder means the thresholds are lower than those in Ref.~\cite{CP13b}. However, this comes with the tradeoff of impressive run times, which enables us to probe very large $d$. It is conjectured in Ref.~\cite{CP13b} that the threshold will follow a constant fraction of the generalized Hashing bound with increasing $d$. However, we see that the obtained thresholds tend away from this limit. This can be partially explained by small system size effects, as we have to decrease the system sizes as we increase $d$.

\section{Summary and Discussion}\label{SecSum}

In this paper we have introduced  two efficient decoding algorithms for decoding the qudit toric code, and we have studied how their thresholds for the independent noise model vary as we change the local dimension of the qudits of the code.

We observed first of all that the HDRG decoder is restricted by the syndrome percolation threshold. For small $d$, the decoder is capable of exploiting the additional syndrome information provided by increasing $d$ and the threshold increased. However, the decoder is unable to correct errors above an error rate of approximately 18\%, where syndromes percolate across the lattice. We introduced an initialization step (in the enhanced-HDRG) which makes use of the charge information given by the syndrome, to annihilate small neutral sub-clusters of syndromes ameliorating the percolation effect. This enhancement enabled the HDRG decoder to achieve thresholds beyond the percolation limit.

Part of the simplicity of the HDRG algorithm is that its minimal utilization of charge information, in particular that the clusterisation step is charge-blind. The initialization step in our enhanced algorithm incorporates more local charge information into the decoding and doing so enhances thresholds. It would be worthwhile to attempt to combine initialization and clusterisation steps into a single algorithm which might combine computational efficiency with higher thresholds.

We study also the SDRG decoder which we optimized for high speed decoding. This decoder uses Bayesian inference methods to coarse grain probability distributions to efficiently find the probabilities that different errors have occurred on the lattice. This decoder considers probabilities of different error configurations at a microscopic level. This enables the decoder to overcome percolation thresholds in a natural way. Instead, we observe that this decoder maintains a threshold which is a constant factor from the optimal threshold for a non-degenerate code, which we should expect to achieve by exhaustive decoding. 

While we see that the SDRG will typically outperform the HDRG decoder, we note that for low $d$ the  HDRG decoder performs comparably well to the SDRG decoder. This is remarkable given the comparative simplicity of the decoder. Moreover, we see that the enhanced-HDRG can continue to achieve thresholds that come close to matching those of the SDRG decoder. % with the small lattice cell size chosen here. 

\begin{figure}
  \centering
    \includegraphics[width=0.5\textwidth]{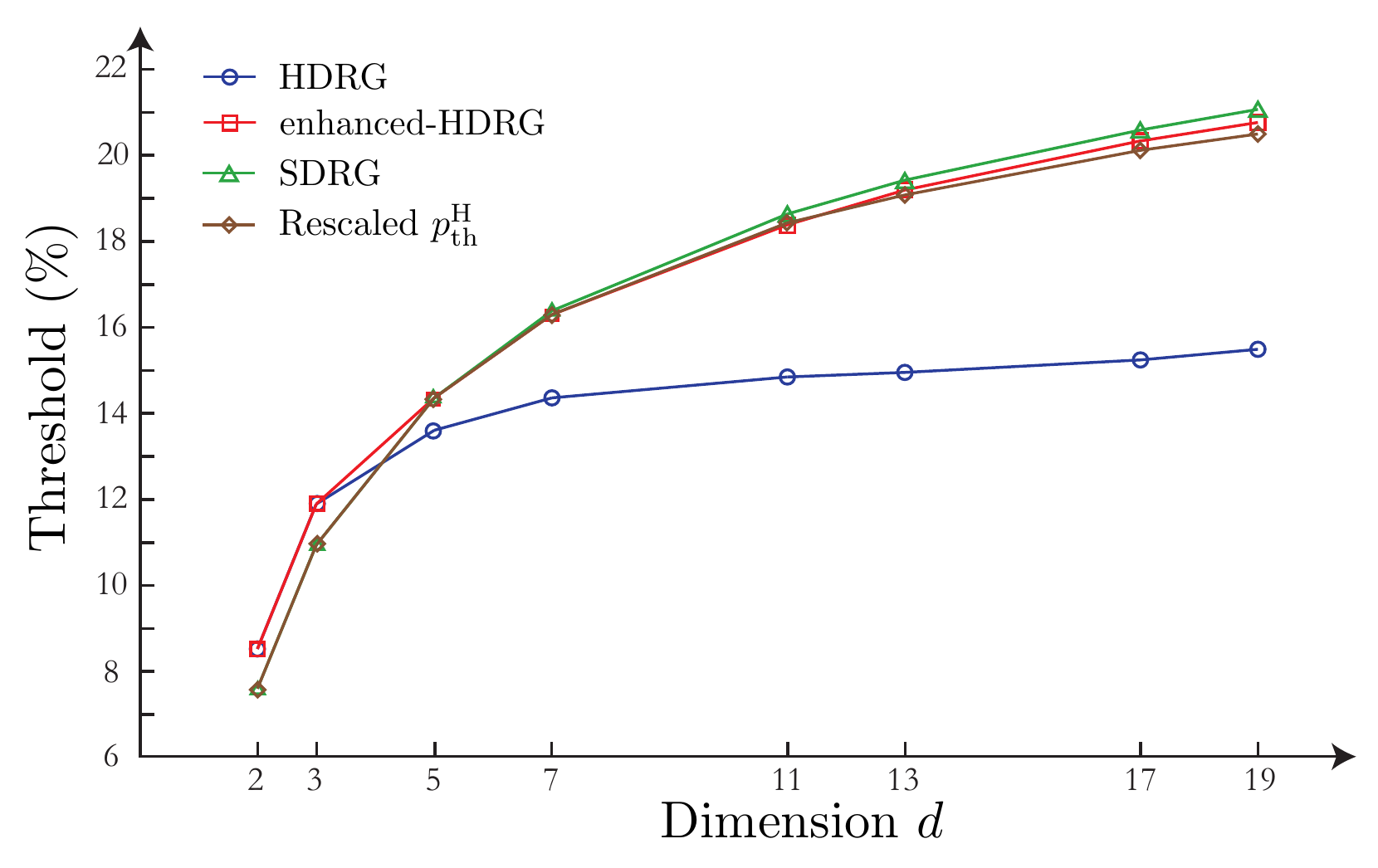}
    \hspace{-5mm}
      \captionsetup{singlelinecheck=yes, justification=centerlast}
      \caption{A comparison between the thresholds obtained using the HDRG, enhanced-HDRG and SDRG decoders presented in this paper, plotted, for comparison, against $0.69 p_{\text{th}}^H$, the hashing bound threshold rescaled to 69\% of its value. In \cite{CP13} Duclos-Cianci and Poulin report that the thresholds for their SDRG decoder (with larger unit cells) are close to $0.81 p_{\text{th}}^H$, independent of $d$.}
      \label{ComparisonHash}
      \vspace{-.5cm}
\end{figure}

We see a general trend of error threshold increasing with dimension $d$. This is in line with the conjecture that the optimal threshold should be close to or equal to the hashing bound threshold. The hashing bound threshold rises monotonically with $d$ up to a maximal value of 50\% for high $d$, and we see a corresponding monotonic rise in the thresholds for both decoders. Comparing the obtained thresholds with a rescaled hashing bound threshold in Fig~\ref{ComparisonHash}, we see further evidence of a phenomenon first reported by Duclos-Cianci and Poulin. For both decoders, the numerical threshold remains close to a constant factor  (69\%) of the hashing bound threshold, independent of $d$. If the conjecture that the hashing bound threshold approximates the optimal threshold is true, this would seem to imply that the decoder thresholds are reaching a constant fraction of the optimal threshold independent of $d$. We do not understand the origin of this effect, and investigating it with a wider range of noise models will be an avenue for future work. 

It is pertinent to discuss some limitations of the noise model studied here. The independent noise model was chosen for its convenience and its connection to statistical mechanics models (the Potts gauge glass). However, it is not physically motivated, and certain aspects of it (equal probability of all powers of $X$ and $Z$, and independence of $X$ and $Z$ errors) do not represent a noise model in nature. In future work, we will explore the performance of these decoders in more general noise models. In particular, for high $d$ one would expect, for general noise channels and e.g. for the depolarizing channel, to see high correlations between $X$ and $Z$ noise. A decoder which took these correlations into account may therefore reach higher thresholds than one treating these as separate decoding problems. It is difficult to make a fair comparison between the noise thresholds of different $d$.  In particular, in the independent noise model we have considered here, as $d$ increases the total error probability is split between more and more individual noise processes. Thus the increased thresholds here must be partly attributed to this fact. Nevertheless, the increase in threshold probability for low $d$ (e.g. from 2 to 3 or 5) is striking and coupled with the increased thresholds and yields observed in magic state distillation at these dimensions \cite{CAB12} may promise advantages in the implementation of quantum computation. However, to verify this promise further study is needed, in particular a full fault-tolerant analysis with a physically motivated noise model allowing fair comparison between schemes of different dimension.

The development of generalized decoding algorithms has provided analytical tools for the study of novel topological systems. For instance, recent developments in decoding algorithms \cite{BH11} have given us a probe to study the properties of topological phases coupled to a thermal environment \cite{BH11,BH13}. In particular, the HDRG decoder developed here is used to study a two-dimensional topological phase with Hamiltonian defects \cite{BAP13}. Moreover, recent advances in decoding algorithms have demonstrated capability to encode and read out quantum information in non-Abelian topological phases \cite{BBD13,WBIL13} which shows promise towards the realization of fault-tolerant topological quantum computation. Further development using more specialized decoders may lead to more refined analysis of such fault-tolerant systems.

Our study shows that both SDRG and HDRG provide effective decoders in scenarios where the MWPMA is inappropriate. The simplicity of the HDRG, and the incorporation of a sub-lattice optimal decoder for the SDRG mean that both may be readily generalized to non-standard topological codes. The key advantage in the HDRG decoder is its light computational requirements. In scenarios where high threshold is important, however, for example, in reducing the code overhead \cite{BV13,fern} is a key priority, the extra classical computational cost of a SDRG decoder, and in particular its ability to reach higher thresholds via larger cell sizes, may be a price worth paying. Overall, the diversity of efficient decoders provides a toolbox for further research into the new phenomena, new physics and potential advantages for quantum information offered by non-traditional and non-qubit topological codes.

\section*{Acknowledgments}

We would like to thank Simon Burton, Guillaume Duclos-Cianci and Fern Watson for useful discussions. We especially thank Fern Watson for her help in analyzing the data and estimating the thresholds.  We acknowledge the Imperial College High Performance Computing Service for computational resources. HA and BJB acknowledge the financial support of the EPSRC (HA is supported by grant number: EP/K022512/1). ETC is supported by the EU (SIQS).

\newpage
\clearpage

\appendix
\section{Homology}\label{HomologyAppendix}

\subsection{Introduction}
The algebra of the stabilizer group of the qudit toric code defined on the edges of a lattice is captured by {\em homology}. Homology is a framework for relating structures of different dimension via the concept of {\em cycles}, which has important applications in topology. We will not give a detailed and formal introduction to homology here, but instead introduce the key concepts needed for understanding homology in toric codes, in terminology accessible to the general physicist. 

In short, homological equivalence of string-like operators supported on the edges of the toric code lattice, as used in the main text and in the literature, correspond to equivalence under multiplication by stabilizer operators---and hence two homologically equivalent operators are logically equivalent on the code-space of the toric code. While this definition will suffice for some readers, we invite those who would like a fuller introduction to homology to read on. For a formal introduction to homology as used in the topological code literature which does not take excursions into more general algebraic topology, we recommend Chap.~3 of Ref.~\cite{Nakahara03} or Chap.~5 of Ref.~\cite{H94}.

\subsection{Simplices and the Triangulation of a Manifold}\label{secsimplex}

\begin{figure}
\includegraphics[scale=2.5]{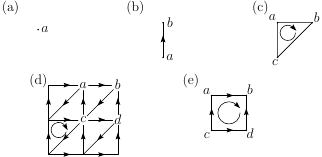}
    	\captionsetup{singlelinecheck=yes, justification=centerlast}
	\caption{\label{homfig0} (a) A 0-simplex, $a$. (b) A 1-simplex. Its orientation is depicted with an arrow from vertex $a$ to vertex $b$. (c) A clockwise oriented 2-simplex. (d) A triangulation of a manifold where 2-simplices have a uniform clockwise orientation. (e) A plaquette, the fundamental square object we use for manifold `triangulation' throughout these notes.}
\end{figure}

The fundamental objects in {\em simplicial homology} which we describe here are {\em directed simplices}. A simplex is an $n$-dimensional generalization of a solid triangle.  A $0$-simplex is a vertex, a $1$-simplex is a line and a $2$-simplex is a triangle. We label vertices with letters. Vertex $a$ is shown in Fig.~\ref{homfig0}(a). The term ``directed'' means we assign orientations to the simplices. The 1-simplex shown in Fig.~\ref{homfig0}(b) is oriented from vertex $a$ to vertex $b$, and the 2-simplex shown in Fig.~\ref{homfig0}(c) has a clockwise orientation from $a$, to $b$, to $c$, and back to vertex $a$.

We introduce the notation $\Delta_{n}$,  to denote an $n$-simplex. We extend this notation to include a direction. A $1$-simplex running from point $a$ to $b$ shown in Fig.~\ref{homfig0}(b), is denoted $\Delta_{1}(a,b)$. The same simplex with opposite direction is written $\Delta_{1}(b,a)$, where we have permuted vertices $a$ and $b$. The 2-simplex shown Fig.~\ref{homfig0}(c) where the clockwise orientation follows the vertices in the sequence $a \rightarrow b \rightarrow c \rightarrow a$, is written $\Delta_{2}(a,b,c)$.  We point out that the sequence of vertices $ b \rightarrow c \rightarrow a \rightarrow b$ will describe the same oriented 2-simplex $\Delta_2(a,b,c)$, such that $\Delta_2(a,b,c) =  \Delta_2(b,c,a) $. The orientation of a 2-simplex is changed by permuting a pair of vertices, for instance, $\Delta_{2}(a,c,b)$ has the opposite orientation to $\Delta_2(a,b,c)$. We should use the notation `$\Delta_0(a)$' to denote a 0-simplex, but for brevity with 0-simplices write only $a$.

Simplices can be used to describe topologically non-trivial manifolds. A manifold, such as the two-dimensional surface of a torus, can be \textit{triangulated}, meaning it can be divided into a set of oriented simplices. We show a triangulation of a two-dimensional manifold in Fig.~\ref{homfig0}(d), where the triangulation includes all the directed 0-, 1- and 2-simplices shown in the diagram. We are free to assign all the 2-simplices of the triangulation a clockwise orientation. 

\subsection{Chains}

\begin{figure}
\includegraphics[scale=2.2]{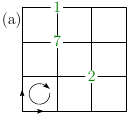}
\includegraphics[scale=2.2]{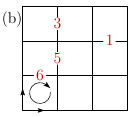}
\includegraphics[scale=2.2]{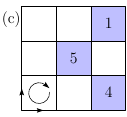}
\captionsetup{singlelinecheck=yes, justification=centerlast}
\caption{\label{homfig1} (a), (b) and (c) show examples of a 0-chain, a 1-chain and a 2-chain respectively on a square lattice. We orient 1- and 2- simplices uniformly. We mark the uniform orientation in the bottom-left corner of the lattice.}
\end{figure}

Having introduced a simplicial triangulation of a manifold, it is now interesting to construct complex objects on a manifold composed of many simplices. General $n$-dimensional objects are known as $n$-{\em chains}. Such $n$-chains are linear combinations of $n$-simplices. We write an $n$-chain, $A$, as 
\begin{equation}
A = \sum_{\Delta} a_{\Delta} \Delta_n,
\end{equation}
where we sum over all $n$-simplices of a triangulated manifold. In the present exposition we consider $a_\Delta \in \mathbb{Z}_d$. We are able to perform binary operations between chains. For example
\begin{equation}
A+B = \sum_{\Delta} (a_\Delta+b_\Delta)\Delta_n,
\end{equation}
where we use $B = \sum_\Delta b_\Delta \Delta_n$. We remark that the additive inverse of a simplex is the same simplex with opposite orientation. For instance, $\Delta_1(a,b) = - \Delta_1(b,a)$, and $\Delta_2(a,b,c) = - \Delta_2(b,a,c)$. 

Having introduced linear combinations of simplices, we are now able to define a {\em plaquette}, $\Xi(a,b,c,d)$, in terms of 2-simplices. The plaquette is the fundamental square object we use to describe the square toric code lattice, shown in Fig.~\ref{homfig0}(e). We consider once more the example triangulation shown in Fig.~\ref{homfig0}(d), we have 
\begin{equation}
\Xi(a,b,c,d) = \Delta_2(a,b,c) + \Delta_2(c,b,d). \label{plaquette}
\end{equation}
We compose an entire lattice of plaquettes. We find the plaquettes of the square decomposition by summing all the pairs 2-simplices which share a {\em diagonal} bounding edge of the considered regularly triangulated manifold. In the next section we see from the example plaquette $\Xi(a,b,c,d)$ that the simplex $\Delta_1(b,c)$ is not included in its bounding set, thus eliminating the diagonal edges of Fig.~\ref{homfig0}(d) from the plaquette decomposition of the manifold.

It is useful to write arbitrary $n$-chains on the square lattice. We will see that such chains correspond to operators relevant to the qudit toric code. We give an example of an arbitrary 0-, 1- and 2-chain on the considered square lattice in Fig.~\ref{homfig1}. In these diagrams, and the diagrams we use throughout the this appendix, we uniformly assign all plaquettes a clockwise orientation, and vertical\,(horizontal) edges are assigned an upwards\,(right) orientation, which we mark in the bottom left corner of each lattice diagram. The numbers then correspond to the coefficient of a given simplex for the described $n$-chain of using the defined orientations.

\subsection{The Boundary Map}
A key idea of homology is the {\em boundary}. A boundary of an $n$-simplex is a unique linear combination of $(n-1)$-simplices. The boundary of $ \Delta_1(a,b)$ shown in Fig.~\ref{homfig0}(b) contains two bounding vertices, $a$ and $b$, and the boundary of a triangle, $ \Delta_2(a,b,c)$, shown in Fig.~\ref{homfig0}(c), contains lines $\Delta_1(a,b)$, $\Delta_1(b,c)$ and $\Delta_1(c,a)$. 

To make the concept of a boundary rigorous, we define the boundary map $\delta_{n}$. The boundary map is a linear map which takes an $n$-chain, $A$, and outputs an $(n-1)$-chain which forms the boundary of $A$. We consider the examples we have introduced in this section.

We first consider a single vertex, $a$, shown in Fig.~\ref{homfig0}(a). A vertex necessarily has no boundary
\begin{equation}
\delta_{0}[a]=0.
\end{equation}
The boundary map $\delta_1$ acting on the 1-simplex $ \Delta_1(a,b)$, shown in Fig~\ref{homfig0}(b) returns
\begin{equation}\label{Eqboundary1}
    \delta_{1} [ \Delta_1(a,b) ] = b-a,
\end{equation}
where the negative sign arises due to the orientation of $\Delta_1(a,b)$. The importance of signs will become clear in later sections where we consider {\em cycles}. 

The last boundary map relevant to us, $ \delta_{2}$, will output a linear combination of the edges. It is defined
\begin{equation}
    \delta_{2}[ \Delta_2(a,b,c) ] = \Delta_1(a,b) - \Delta_1(a,c) + \Delta_1(b,c), \label{2boundarymap}
\end{equation}
again, where it is very important to keep track of vertex order $a$, $b$ and $c$ to maintain consistency with the signs. One can easily check that the output $\delta_2[\Delta_2(a,b,c)]$ is independent of even (cyclic) permutations of vertices $a$, $b$ and $c$ using that $\Delta_1(a,b) = - \Delta_1(b,a)$. 

Finally, we consider the boundary of a plaquette~(\ref{plaquette}), composed from 1-simplices which we denote $\partial p$. By linearity we have that
\begin{eqnarray}
  \partial p \equiv  \delta_2[  \Xi (a,b,c,d) ] &=& \delta_2[ \Delta_{2}(a,b,c) ]  \nonumber \\ 
  && + \delta_2 [  \Delta_{2}(b,d,c) ].
\end{eqnarray}
Then, using Eqn.~(\ref{2boundarymap}), it is easily checked that
\begin{equation}
\label{boundPlaq}
\partial p= \Delta_{1}(a,b) + \Delta_{1}(c,a)   - \Delta_{1}(d,b)- \Delta_{1}(c,d).
\end{equation}
As the orientations of the two simplices of $\Xi(a,b,c,d)$ align, the boundary matches with our intuition and the $\Delta_1(b,c)$ does not appear in the boundary of the plaquette.

\subsection{Cycles}

We begin discussing cycles by considering the example of the boundary of a plaquette, calculated in Eqn.~(\ref{boundPlaq}). We calculate the boundary of this 1-chain using Eqn.~(\ref{Eqboundary1}) to find
\begin{eqnarray}    
\delta_{1} [\partial p] &=& (b-a) -(c-a) +(d-b) - (d-c)   = 0. \label{BoundaryResult}
\end{eqnarray}
In homology any $n$-chain, $A$, such that $\delta_n[A] = 0$, is known as an $n$-{\em cycle}. Calculation~(\ref{BoundaryResult}) shows explicitly that the boundary of a plaquette is a $1$-cycle. Indeed, one can verify in general that a boundary of any $n$-chain is a cycle. Written more rigorously, one can prove that $\delta_{n-1}[ \delta_{n} [ A ] ]=0$ for any $n$-chain $A$.

\begin{figure}
\begin{center}
\includegraphics[width=0.4\columnwidth]{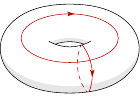}
\end{center}
\captionsetup{singlelinecheck=yes, justification=centerlast}
\caption{\label{homfig5} On a torus there are a number of ways one can construct cycles which do not enclose a boundary. Here are two examples. The cycles depicted here cannot be transformed to one another via the addition of homologically trivial cycle. 
 }
\end{figure}

Are all $n$-cycles the boundary of an $(n+1)$-chain?  The answer is no. Consider the loop indicated in Fig.~\ref{homfig5}. This chain has zero boundary, which can be verified algebraically. Nevertheless it encloses no 2-chain. If we try and ``fill out'' the surface of the torus to enclose a chain, we will find that this covers the whole toric surface. The $2$-chain covering the surface however, has no boundary. Hence this $1$-cycle is not a boundary of any $2$-chain.

The distinction between boundary cycles and non-boundary cycles is of central importance to homology theory (and to the toric code). Boundary cycles are known as ``homologically trivial'' cycles. Otherwise, cycles are ``homologically non-trivial''.

\subsection{Homological Equivalence}

\begin{figure}[h!]
\begin{center}
\includegraphics[scale=2.2]{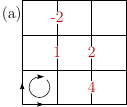} 
\includegraphics[scale=2.2]{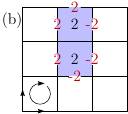} 
\includegraphics[scale=2.2]{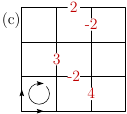}
\end{center}
	\vspace{-.3cm}
	\captionsetup{singlelinecheck=yes, justification=centerlast}
	\caption{\label{homfig6} An example of homology. (a) Shows a 1-chain $A$ (b) A 1-cycle $\partial H$ which bounds 2-chain $H$ which is highlighted on blue plaquettes. (c) A 1-chain $B$ homologous to $A$ as $B = A+ \partial H$. }
\end{figure}

The group of boundary cycles is used to define another central concept of homology theory, {\em homological equivalence}. We say two $n$-chains are homologically equivalent, or {\em homologous}, if they are equivalent up to addition of a boundary $n$-cycle. We provide an explicit example of two homologous 1-chains. We show in Fig.~\ref{homfig6}(a) and Fig.~\ref{homfig6}(c) the 1-chains $A$ and $B$ respectively. It's easily seen that the $B$ differs from $A$ by only a boundary, $\partial H$, shown in Fig.~\ref{homfig6}(b). The 1-cycle $\partial H = \delta_2[H]$ is clearly a boundary of the 2-chain $h$ highlighted in blue on Fig.~\ref{homfig6}(b). We denote that two $n$-chains are homologous by the symbol `$\sim$', such that $ A \sim B $. It is from this insight that we see why boundary $n$-cycles are known as ``homologically trivial''. All $n$-boundaries are homologous to the trivial $n$-chain, $A = 0$. It is in this sense that homologically trivial cycles are {\em contractible}, i.e. can be contracted to a point or the trivial $n$-chain. Non-trivial cycles such as those shown in red in Fig.~\ref{homfig5} do not share this property. Indeed, these non-trivial cycles are known as {\em non contractible}.

Before we move onto the final section of this appendix we remark that the group of non-trivial cycles of a triangulation of a manifold is known as the {\em first homology group}, and is a topological invariant used to classify manifolds.

\subsection{Homology and the Toric Code}

Here we arrive at the main section of the appendix where we show that operators acting on the code-space of the qudit toric code are elegantly characterized using concepts from homology. 

We make use of the notation introduced in the main text to identify 1-chains $A = \sum_\Delta a_\Delta \Delta$ with Pauli operators $Z(A) = \bigotimes_\Delta Z_{\Delta}^{a_\Delta}$, The subscript $\Delta$ now indexes qudits lying on edges of the toric code lattice.

\begin{figure}
\begin{center}
\vspace{.5cm}
\includegraphics[scale=2.5]{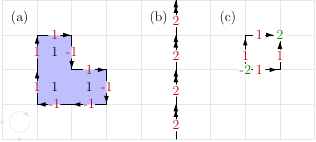}
\captionsetup{singlelinecheck=yes, justification=centerlast}
\caption{\label{homfig8} (a) A homologically trivial 1-cycle which is the boundary of the 2-chain highlighted in blue. (b) A homologically non-trivial 1-cycle $C$.  (c) A 1-chain whose boundary is non-zero.}
\end{center}
\end{figure}

We first consider the plaquette operators of the toric code. It is easily checked that plaquette stabilizers simply correspond to the boundary cycle of a plaquette, such that $B_p = Z(\partial p)$, where $ \partial p = \delta_1[\Xi(a,b,c,d)]$. In fact, it is easily checked that any operator of the form $Z(\partial A)$ where $\partial A = \delta_2(A)$ for any 2-chain $A$ will act trivially on the code-space of the toric code. We show an example of such a boundary cycle in Fig.~\ref{homfig8}(a).

We consider next logical $\bar{Z}$ operators of the qudit toric code. These are easily identified with homologically non-trivial 1-cycles, such as $C$, shown in Fig.~\ref{homfig8}(b). A sensible encoding of the toric code might be chosen such that $\bar{Z}^2 = Z(C)$. All operators $Z(C')$ where $C' \sim C$ will act equivalently on the code space of the qudit toric code to the operator $Z(C)$.

We have seen in this subsection that the code-space of the toric code is acted on by operators of form $Z(C)$, where trivial cycles $C$ act trivially on the code-space and non-trivial cycles $C$ perform logical operations on the code-space. In fact, the vertex operators are prescribed such that the syndromes of operators $Z(C)$ for 1-chains $C$ which are not cycles will introduce syndromes equal to the boundary of the 1-chain, $\delta_1[C]$. We see an example of a 1-chain with its corresponding boundary written in green in Fig.~\ref{homfig8}(c). Once more it is easily checked that chains homologous to $C$ will generate the same syndrome, by simple calculation, we find the boundary of $C' = C + \partial A$ where $\partial A$ is the boundary of a 2-chain
\begin{eqnarray*}
\delta_0[C' ] &=&  \delta_0[C + \partial A ], \\
& = & \delta_0[C] + \delta_0[\partial A] = \delta_0[C]. 
\end{eqnarray*}

Finally, we remark that all the homological properties of vertex stabilizers, logical $\bar{X}$ operators and $X$-type error chains are the same if we move to a dual lattice. On the dual lattice, the role of plaquettes and vertex operators are interchanged, the same homology mapping captures the relationship between $X$-errors, logical $\bar{X}$ operators, plaquette syndromes. We summaries the correspondences between the toric code properties and homology concepts in Tab.~\ref{homologytable}.

\begin{table*}
\centering
\begin{tabular}{|l | c |  l| }
\hline
Toric Code Property & Lattice &  Homological Description\\\hline
Plaquette ($Z$) stabilizer subgroup & Primal & Set of homologically trivial 1-cycles\\
Vertex ($X$) stabilizer subgroup & Dual & Set of homologically trivial 1-cycles\\
$Z^k$ error configuration & Primal & 1-chain\\
Vertex syndrome configuration & Primal & Boundary of $Z$-error 1-chain\\
$X^k$ error configuration & Dual & 1-chain\\
Plaquette syndrome configuration & Dual & Boundary of $X$-error 1-chain\\
$\bar{Z}^k$ logical operator & Primal & Homologically non-trivial 1-cycle\\
$\bar{X}^k$ logical operator & Dual & Homologically non-trivial 1-cycle\\\hline
\end{tabular}

\caption{\label{homologytable} A table showing the relationship between $X$- and $Z$-type operators on the qudit toric code and their respective chain in homology theory.}
\end{table*}

\section{Hashing Bound Threshold}\label{AppSecHash}
 
The hashing bound is an important quantity from quantum Shannon theory \cite{W13book}. It is often described in relation to the capacity of a communication channel. For instance, consider the Pauli noise channel: A channel with Kraus operators $\sqrt{p_j}\sigma_j$ where $\sigma_j$ are the (qubit or qudit) Pauli operators (including the identity) and $\sum_{j}{p_j}=1$. Then the hashing bound represents a lower bound on the capacity of this channel \cite{W13book,BDS96}. This bound is given by the rate $R$ achievable by using a random coding protocol, given by
\begin{equation}\label{hashingbound}
R = 1 - H(p_j),
\end{equation}
where $H$ is the base-$2$ entropy defined as
\begin{equation}\label{entropy}
H(p)=-\sum_j p_j \log_2(p_j).
\end{equation}

We shall call the values of $p_j$ at which $R$ reaches zero is the {\em hashing bound threshold}, denoted here as $ p_{\mathrm{th}}^{\mathrm{H}} $. Note that some authors call the hashing bound threshold simply the hashing bound.

For one parameter noise families, the hashing bound threshold is given by a single value of that parameter. For example, for the qubit independent noise model, where $X$ and $Z$ errors occur independently with probability $p$, i.e. $p_x= p_z= p(1-p)$, $p_y= p^2$ and $p_\mathbb{1}=(1-p)^2$, the hashing bound threshold is  $p_{\mathrm{th}}^{\mathrm{H}}=0.110028\%$ (to 6 d.p.). The closeness between this value and the optimal threshold for the qubit toric code under the independent noise model was noted by Dennis \textit{et al.} \cite{DKL02}. 

Dennis \textit{et al.} also showed that the optimal decoder for the qubit toric code can be mapped to the Random-Bond Ising Model (RBIM) with the optimal threshold corresponding to  a phase transition point known as the Nishimori point. The generalization of this mapping to the qudit toric code of their argument is straight-forward and leads to a model known as the Potts gauge glass (PGG). 

Further work by Nishimori and collaborators \cite{NN02,TN05} implied that the similarity between the optimal and the hashing bound thresholds applies to more general statistical mechanical models. In particular they showed the the Nishimori point for the RBIM and PGG could be estimated via a duality argument. The value of the  Nishimori point (and thus the optimal decoder threshold) they derive is identical to the hashing bound threshold for the independent noise model.

A similar close relationship between the hashing bound threshold and optimal threshold for different noise models has also been observed. For example, the optimal threshold of the qubit toric code for depolarizing noise is estimated to be  $ p_{\mathrm{th}}^{\mathrm{opt}}=18.9\% $ \cite{BAO12}, and this, again, is very close to the hashing bound for that noise model $ p_{\mathrm{th}}^{\mathrm{H}}=18.93\% $.

\begin{figure}
\includegraphics[width=8cm]{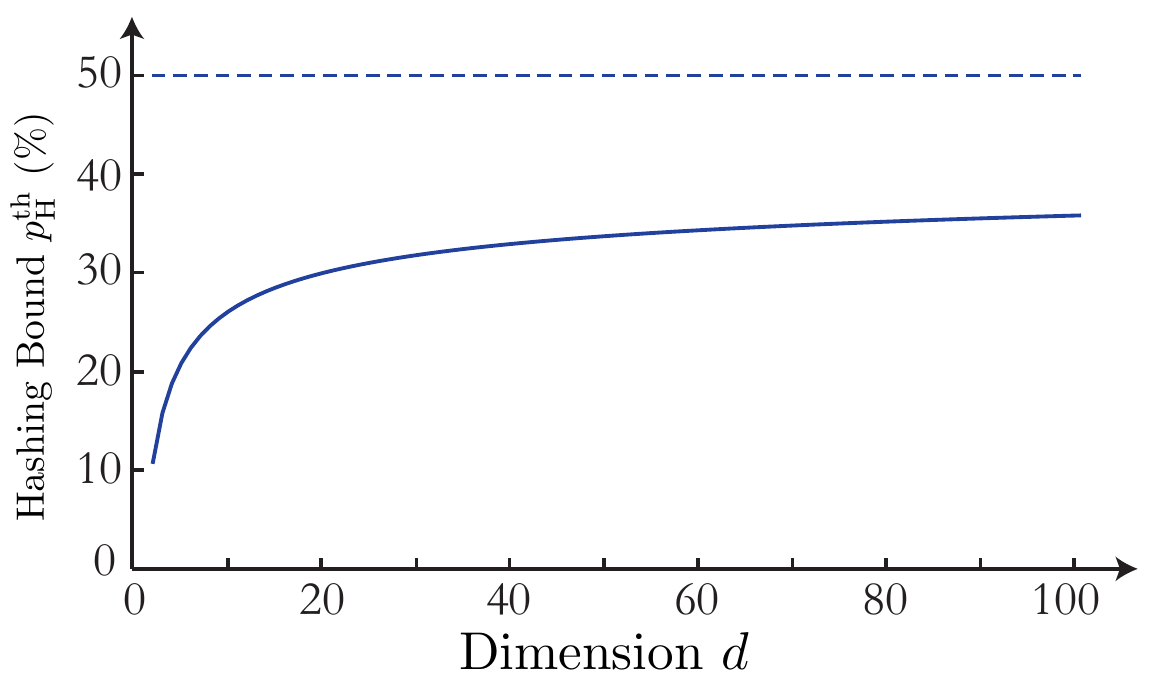}
\caption{\label{hashingfig} The hashing bound threshold for the independent error model (defined in the main text) for parameter $p$ as a function of qudit dimension $d$.}
\end{figure}

Given the evidence that the hashing bound threshold is close to the optimal decoder threshold for qudit codes under the independent noise model, it represents a natural point of comparison for the thresholds in our study. We plot in Fig.~\ref{hashingfig} the hashing bound thresholds for this error model as a function of dimension $d$. In the limit of $ d\rightarrow\infty $ the hashing bound threshold $ p_{\mathrm{th}}^{H}\rightarrow 50\% $.

\end{document}